\documentclass[10pt,journal]{IEEEtran}
\usepackage{amsmath,amsfonts}
\usepackage{algorithmic}
\usepackage{algorithm}
\usepackage{array}
\usepackage[caption=false,font=normalsize,labelfont=sf,textfont=sf]{subfig}
\usepackage{textcomp}
\usepackage{stfloats}
\usepackage{url}
\usepackage{verbatim}
\usepackage{graphicx}
\usepackage{cite}
\usepackage{booktabs}
\usepackage{makecell}
\usepackage{xcolor}

\begin{document}

\title{Time Synchronization for 5G and TSN\\ Integrated Networking}

\author{Zixiao~Wang, Zonghui~Li, Xuan~Qiao, Yiming~Zheng, Bo~Ai~\IEEEmembership{Fellow,~IEEE,} and Xiaoyu~Song

    \thanks{Zonghui Li, Zixiao Wang, and Xuan Qiao are with the School of Computer and Information Technology, Beijing Jiaotong University, Beijing, China, 100044. Zonghui Li is the corresponding author. E-mail: zonghui.lee@gmail.com, w\_zixiao@bjtu.edu.cn, 21120386@bjtu.edu.cn.}
    \thanks{Yiming Zheng, Bo Ai are with the School of Electronic and Information Engineering, Beijing Jiaotong University, Beijing, China, 100044. E-mail: 23111055@bjtu.edu.cn, boai@bjtu.edu.cn.}
    \thanks{Xiaoyu Song is in the Department of Electrical and Computer Engineering, Portland State University, Portland, OR. E-mail: songx@pdx.edu.}
}



\maketitle

\begin{abstract}
Emerging industrial applications involving robotic collaborative operations and mobile robots require a more reliable and precise wireless network for deterministic data transmission. To meet this demand, the 3rd Generation Partnership Project (3GPP) is promoting the integration of 5th Generation Mobile Communication Technology (5G) and Time-Sensitive Networking (TSN). Time synchronization is essential for deterministic data transmission. Based on the 3GPP's vision of the 5G and TSN integrated networking with interoperability, we improve the time synchronization of TSN to conquer the multi-gNB competition, re-transmission, and mobility problems for the integrated 5G time synchronization. We implemented the improvement mechanisms and systematically validated the performance of 5G+TSN time synchronization. Based on the simulation in 500m x 500m industrial environments, the improved time synchronization achieved a precision of 1 microsecond with interoperability between 5G nodes and TSN nodes.
\end{abstract}

\begin{IEEEkeywords}
TSN, 5G, time synchronization, IEEE 1588.
\end{IEEEkeywords}

\section{Introduction}
\IEEEPARstart{T}{he} emergence of Industry 4.0 has made information and data play an increasingly important role in the development of industrial manufacturing and information and communication technology \cite{Hermann2016D}. Industry 4.0 is guiding the industrial field toward more intelligence, flexibility, and efficiency. As the main carrier of data transmission, the network plays a crucial role in the industrial field. The characteristics of Industry 4.0 are driving the increasing demand for network universality in the field of industrial control \cite{Vit2019I}.

Traditional industrial control networks excel in ensuring high reliability in industrial control domains, but they also suffer from limitations in network bandwidth and relatively fixed topology. Therefore, Industrial Ethernet provides a new solution to meet the requirements of Industry 4.0, aiming to compensate for the shortcomings of traditional bus-type industrial control networks in terms of bandwidth, speed, and scalability \cite{Bru2019A}. In 2012, the IEEE 802.1 initiated a Time-Sensitive Networking (TSN) working group to standardize the development of industrial Ethernet. This working group is responsible for extending standard Ethernet to support real-time and deterministic data transmission. Tab.\ref{tab:TSN characteristics} enumerates the characteristics of TSN.
\begin{table}[!t]
\caption{the Characteristics of Time-Sensitive Networking}
    \centering
    \begin{tabular}{l l l}
        \toprule
        \textbf{service} & \multicolumn{2}{l}{\textbf{advantage}} \\
        \midrule
        Delay and jitter & \multicolumn{2}{l}{\makecell[l]{time synchronization precision lass than 1$\mu$s, \\ data transmission jitter lass than 5$\mu$s.}} \\
        bandwidth & \multicolumn{2}{l}{\makecell[l]{based on full duplex Ethernet 1Gbps \\ and 10Gbps networks.}} \\
        reliability & \multicolumn{2}{l}{\makecell[l]{Data frame replication and elimination; \\ Path redundancy.}} \\
        Interoperability & \multicolumn{2}{l}{\makecell[l]{Based on standard Ethernet development; \\ Adopting centralized management.}} \\
        \bottomrule
    \end{tabular}
    \label{tab:TSN characteristics}
\end{table}

TSN provides low latency, high reliability, and real-time communication services while remaining compatible with standard Ethernet. This compatibility allows TSN and standard Ethernet data streams to coexist and be transmitted over the network \cite{ISO2019ISO}. Real-time capabilities and determinism are important features of TSN. They ensure that TSN packets are transmitted within well-defined network latency boundaries and the jitter of transmission latency within the range of microseconds or lower \cite{Lo2019A}.

Meanwhile, the application of mobile robots in fields such as collaborative transportation is steadily increasing. These applications require high-performance wireless industrial internet communication systems. The current wireless communication technologies for the Industrial Internet of Things (IIoT), such as IO-Link Wireless over Bluetooth and industrial wireless LAN by Siemens, are primarily suitable for localized IIoT applications with low data rates or low mobility. These technologies cannot meet the requirements of existing industrial use cases for high traffic, high speed, and high mobility \cite{Mah2022I}. The 5th Generation Mobile Communication Technology (5G) with various advanced wireless technologies, unified connectivity, and wide coverage capabilities has more advantages than other solutions \cite{Vit2019I}. The 3rd Generation Partnership Project (3GPP) has promoted the support of TSN in 5G standard Release-16 and Release-17 \cite{3GPP2021S}. One significant challenge to achieve deterministic interoperability in the integrated networking of 5G and TSN is establishing the 5G+TSN high-precise time synchronization by conquering the multi-gNB competition, re-transmission, and mobility problems of 5G. This paper makes the following contributions to the time synchronization in 5G and TSN integrated networking:

\begin{itemize}
    \item Improving and deploying IEEE 1588 and its variant IEEE 802.1AS in 5G to meet industrial mobile applications' operational high-precise time synchronization requirements (1 microsecond).
    \item Validate time synchronization performance within integrated 5G and TSN networking and study its application in industrial environments.
\end{itemize}

The paper is organized as follows. Sec.~\ref{sec:Related Work} surveys the related work of time synchronization on 5G and TSN. Sec.~\ref{sec:background} presents the time-synchronization background of 5G+TSN. Sec.~\ref{sec:design} details the improved designs for time synchronization in the 5G and TSN integrated networking. Sec.~\ref{sec:test and validate} firstly validates the synchronization performance of the improvement mechanism. Subsequently, the whole 5G+TSN time synchronization is evaluated in the OMNeT++ simulation platform, and a case study in an industrial application environment is performed. Finally, Sec.~\ref{sec:conclusion} concludes the paper.

\section{Related Work} \label{sec:Related Work}
In the 5G standard Release-16, 3GPP introduced a time-sensitive communication architecture for the 5G System to expand their support for TSN \cite{Par20205G, Pei20205G}. This signifies the acceleration of forming a 5G and TSN closed-loop industrial control network system, transitioning from wired to wireless, with the backbone network employing TSN and the edge utilizing 5G \cite{Nas2019U}.

\begin{figure*}[!t]
    \centering
    \includegraphics[width=0.95\linewidth]{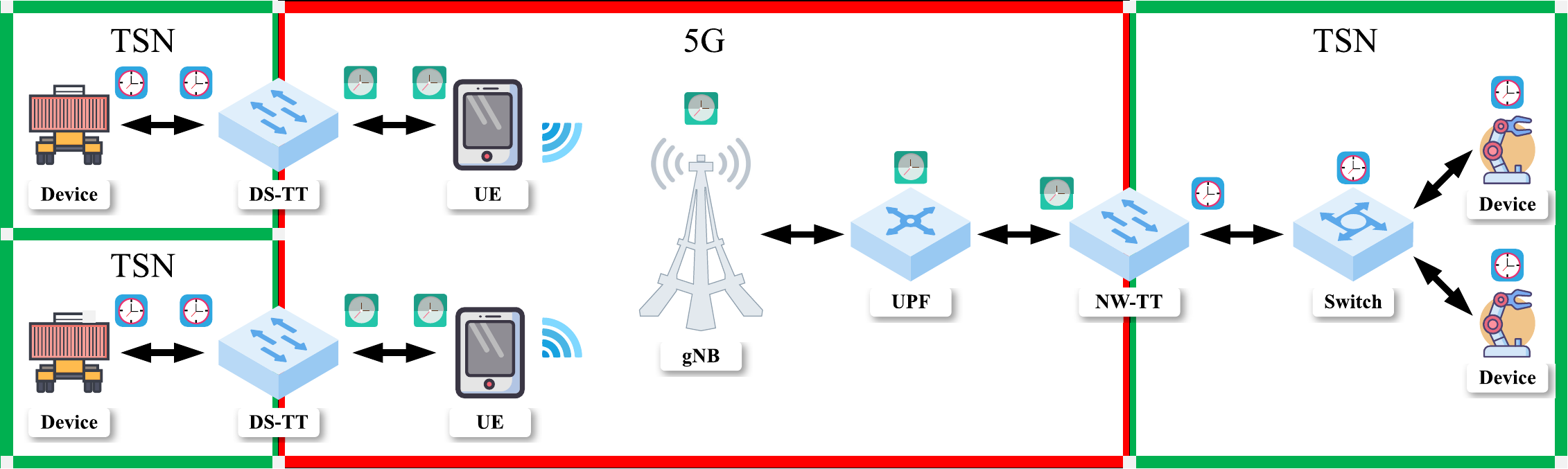}
    \caption{\textcolor{black}{The 5G and TSN Integrated Networking architecture proposed by 3GPP. the TSN network is connected to the device-side TSN network through a 5G transparent bridge. The 5G Core Network is wired to the Network-Side TSN network, and the Device-Side TSN network is wired to the 5G terminal. The Device-Side TSN network realizes the connection with the Network-Side TSN network through the NG-RAN.}}
    \label{fig:Old_Framework}
\end{figure*}

3GPP proposed a low-cost integrated networking for 5G+TSN, which presents a 5G system as a TSN transparent bridge (5G transparent bridge), just like any other TSN transparent bridge \cite{3GPP2021Sys}. The 5G system connects to TSN through the Network-Side TSN Translator (NW-TT) and the Device-Side TSN Translator (DS-TT), as shown in Fig.~\ref{fig:Old_Framework}. \cite{Sat2022D} gave the simulation implementation of NW-TT and DS-TT as the transparent bridge. \cite{Gun2020I} evaluated the impacts on time synchronization of TSN after going across a 5G transparent bridge in a small and ideal discrete testbed. Furthermore, \cite{Gar2022A} explores the long-distance impacts on time synchronization by enlarging the 5G transparent bridge. Since the 5G system is treated as a transparent bridge, when a sync frame goes across the 5G system, the sync frame only records its resident time in the 5G system and does not synchronize the time of the 5G system. As a result, \cite{Wang2023R, Li2023D, Mart2020S} demonstrated the millisecond-level jitter when frames went from 5G to TSN because frames missed the scheduled sending time points of TSN due to un-synchronized time between 5G and TSN.  

To enhance the deterministic transmission in the integrated 5G and TSN networking with interoperability, 3GPP in Release-17 \cite{3GPP2021S} supports time synchronization between 5G and TSN instead of the 5G transparent bridge in Release-16. It is indicated that when the Application Function (AF) and the 5G system are in different trust domains, the AF can utilize the Time Sensitive Communication Time Synchronization Function (TSCTSF) within the 5G system by Network Exposure Function (NEF). When the AF and 5G systems are in the same trust domain, TSCTSF can be used directly. To support time synchronization by TSN, Release-17 defines that a 5G system can operate in any mode within the IEEE 802.1AS Time-Aware Domain, IEEE 1588 Boundary Clock, IEEE 1588 Peer-to-Peer Transparent Clock, and IEEE 1588 End-to-End Transparent Clock. Besides, in Release-17, the concept of non-public networks (NPNs) has been introduced. NPNs allow users to deploy private 5G mobile networks within closed operational environments. Compared to public networks, private networks offer unparalleled advantages in terms of performance, privacy, and security. These enhancements in Release-17 provide the architectural foundation for 5G and TSN integrated networking to achieve time synchronization between TSN and 5G.

\cite{Striffler2019T, Sat2022D} presents the advantages of end-to-end deterministic transmission by sharing the time information between 5G and TSN. However, using IEEE 1588 and its variant 802.1AS of wired Ethernet directly to synchronize wireless 5G is hard. To our knowledge, this paper is the first to systematically implement the time synchronization between 5G and TSN by conquering the multi-gNB competition, re-transmission, and mobility problems of 5G.

\section{Background} \label{sec:background}
\subsection{Time Synchronization In TSN}
\subsubsection{Transparent Clock Mechanism}
\begin{figure}[!t]
    \centering
    \includegraphics[width=0.9\columnwidth]{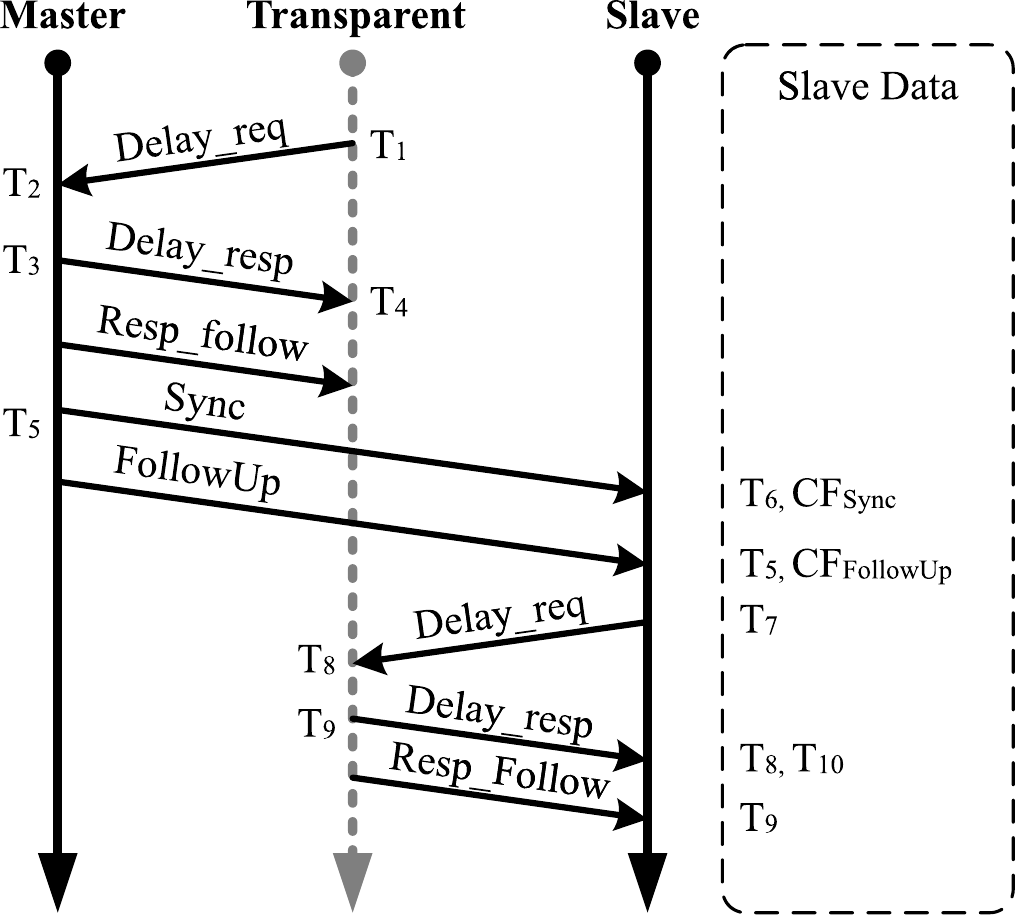}
    \caption{\textcolor{black}{Operation flow of time synchronisation protocols in IEEE 1588 and 802.1AS protocols, implemented using a Peer-to-Peer Transparent Clock Mechanism.}}
    \label{fig:P2PTransparentClock}
\end{figure}

IEEE 802.1AS is a widely used time synchronization mechanism in TSN \cite{Ker2019P}. The bridge nodes in IEEE 802.1AS adopt a similar implementation as the Peer-to-Peer Transparent Clock in IEEE 1588. Peer-to-Peer Transparent Clock is based on the Peer Delay Measurement mechanism. When a Peer-To-Peer Transparent Clock forwards a $\textit{Sync}$ message, it records the peer delay between itself and its master node in the Correction Field (CF). The working process is illustrated in Fig.~\ref{fig:P2PTransparentClock}. In Peer-To-Peer Transparent Clock mode, the Slave Clock can directly calculate the time offset without requesting link delay after receiving the synchronization message:
\begin{align}
    \textit{PeerDelay}_{base}=\frac{\left( T_4-T_1 \right) -\left( T_3-T_2 \right)}{2}
\end{align}

\subsubsection{Frequency Synchronization}
\begin{figure}[!t]
    \centering
    \includegraphics[width=0.9\columnwidth]{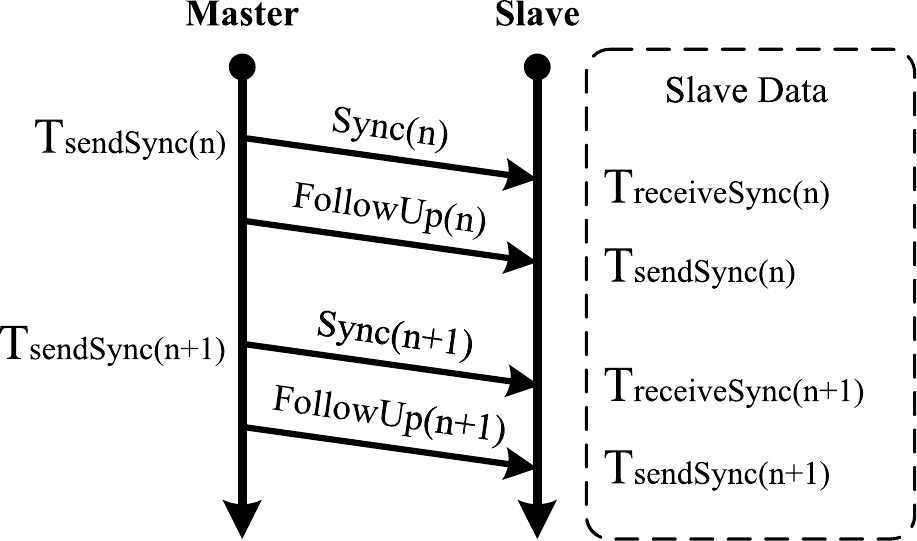}
    \caption{Clock Frequency Synchronization Mechanism. \textcolor{black}{It is used to calculate the difference in clock frequency between master and slave nodes from the time synchronization information of two adjacent times.}}
    \label{fig:frequency_synchronism}
\end{figure}

Clock frequency refers to the number of oscillations per second within a crystal oscillator clock. When the clock frequencies of the master and slave clocks are different, it can affect timestamp precision for time synchronization messages, reducing synchronization precision. The use of a frequency synchronization mechanism can reduce the impact on timestamps. IEEE 1588 provides a frequency synchronization algorithm, as shown in Fig.~\ref{fig:frequency_synchronism}.

In Fig.~\ref{fig:frequency_synchronism}, $T_{sendSync}$ is the sent timestamp of the $\textit{Sync}$ message and $T_{receiveSync}$ is the received timestamp of the $\textit{Sync}$ message. The clock frequency ratio is calculated using timestamps from twice continuous time synchronization.
\begin{align}
    \textit{rateRatio}=\frac{T_{\textit{sendSync}\left( n+1 \right)}-T_{\textit{sendSync}\left( n \right)}}{T_{\textit{receiveSync}\left( n+1 \right)}-T_{\textit{receiveSync}\left( n \right)}}
\end{align}

\subsubsection{Time Synchronization}
In the time synchronization, the Master Node within the synchronization domain periodically sends $\textit{Sync}$ and $\textit{Follow Up}$ to the slave nodes according to the configured sync interval. Subsequently, slave nodes periodically request peer delay measurement according to the peer delay measurement interval.

In Fig.~\ref{fig:P2PTransparentClock}, when the child node sends $\textit{Delay\_req}$ to its parent node, and the parent node responds with $\textit{Delay\_resp}$ and $\textit{Resp\_Follow}$, peer delay is computed. And time synchronization offset is computed when the child node successfully receives $\textit{Sync}$ and $\textit{Follow Up}$.
\begin{align} 
    \textit{PeerDelay}&=\frac{\textit{rateRatio}\times \left( T_4-T_1 \right) -\left( T_3-T_2 \right) }{2} \label{eq:peer delay measurement} \\
    CF &=CF_{Sync} + CF_{Follow Up} \\
    \textit{offset}&=\text{T}_{6}-\text{T}_{5}-\textit{PeerDelay}-CF \label{eq:sync}
\end{align}

\subsection{Time Synchronization For 5G}
In the 5G System, synchronization can be distinguished into two parts: 5G Core Network and Next Generation Radio Access Network (NG-RAN). In the 5G Core Network, there are various ways to achieve time synchronization between nodes, such as the Global Positioning System (GPS), Compass Navigation Satellite System (CNSS), and IEEE 1588. However, GPS and CNSS are not always online and stable due to different physical positions and climates. IEEE 1588 is preferred. NG-RAN defines a Synchronization Signal Block (SSB) for synchronization in the time-frequency domain. gNB transmits time-frequency domain information to UE by sending Primary Synchronization Signal (PSS) and Secondary Synchronization Signal (SSS) in SSB \cite{omri2019S}. This is the foundation for UE to properly receive and transmit data, enabling time synchronization between UEs \cite{Korde2020S}. However, the synchronization mechanism of NG-RAN runs at the physical layer and cannot share time information with the 5G Core Network and TSN without changing the current design. To achieve a fully synchronized 5G and TSN integrated networking, we implemented IEEE 1588 based on UDP/IPv4 in the NG-RAN \cite{IEEE2009P}, which unifies the usage of IEEE 1588 for time synchronization in 5G system including Core Network and NG-RAN.

\section{Design For 5G + TSN Time Synchronization} \label{sec:design}
In this section, we systematically present the solution for 5G+TSN time synchronization. Since IEEE 1588 is well-established in wired Ethernet. So, we mainly present the improved designs for IEEE 1588 to achieve high-precise time synchronization for 5G systems. That is to conquer the multi-gNB competition, re-transmission, and mobility problems of 5G.

\subsection{Obtain High-Precision Timestamps in 5G Network}
According to IEEE 1588 PTP Protocol, the detection precision of timestamps is a crucial factor that enhances time synchronization precision compared to Network Time Protocol(NTP). The closer the timestamp detection module is to the physical layer, the less influence the residence time of time synchronization messages in the network protocol stack have on synchronization precision. However, getting timestamps directly from physical layer interfaces is impossible under the current 5G NG-RAN architecture.

To solve the problem of deploying IEEE 1588 in Wi-Fi networks, \cite{Chen2023U} proposed placing the timestamp detection module in the interrupt handler of the Wi-Fi Network Interface Card (WNIC). When the device completes receiving or transmitting events, it triggers an interrupt event and reports the timestamp to the high-layer PTP program. We have implemented this approach in 5G networks. We modified the network interface driver to obtain the send or receive hardware timestamps within the interrupt handlers. The timestamp detection module is positioned between the data link and physical layers in such a model.

\subsection{Handling Sync Error in gNBs Competition} \label{sec:Dual Clock Counter Mode}
\begin{figure}[!t]
    \centering
    \includegraphics[width=1\columnwidth]{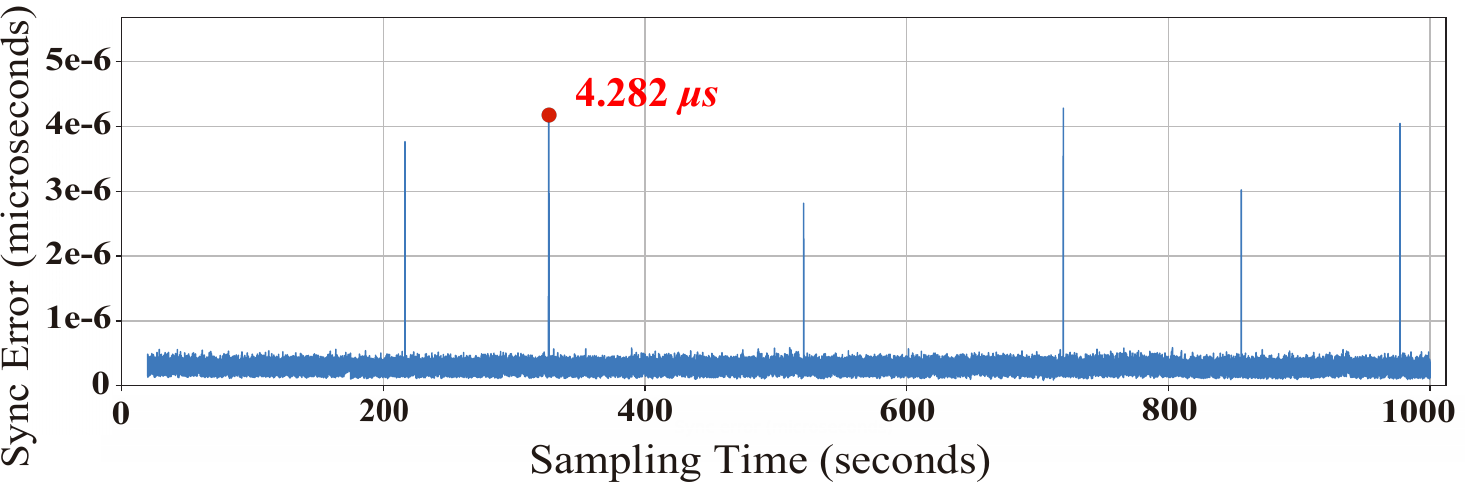}
    \caption{Sampling Result for Time Synchronization \textcolor{black}{when using the same clock counter for generating message timestamps and applying time synchronization offset. $4.282\mu s$ is the maximum value of sync errors in the results, and each sync error persisted for some time before being eliminated}. }
    \label{fig:Sync Crash result}
\end{figure}

Without considering NG-RAN, we conducted a preliminary stability test of the time synchronization mechanism between the 5G Core Network and TSN in OMNeT++. During the simulation, the time synchronization program runs in a network with 100 TSN terminals and 100 Next-generation NodeBs (gNBs) for 1000 seconds. According to the sampling result in Fig.~\ref{fig:Sync Crash result}, it was observed that at certain moments, each synchronization error lasts for several seconds at a time. There were significant fluctuations in the time synchronization result. We analyzed time synchronization and peer delay measurement results in detail. We have discovered two synchronization issues that we have named "Synchronization Collision".

\subsubsection{Synchronization Collision Scenario 1}
\begin{figure}[!t]
    \centering
    \includegraphics[width=\columnwidth]{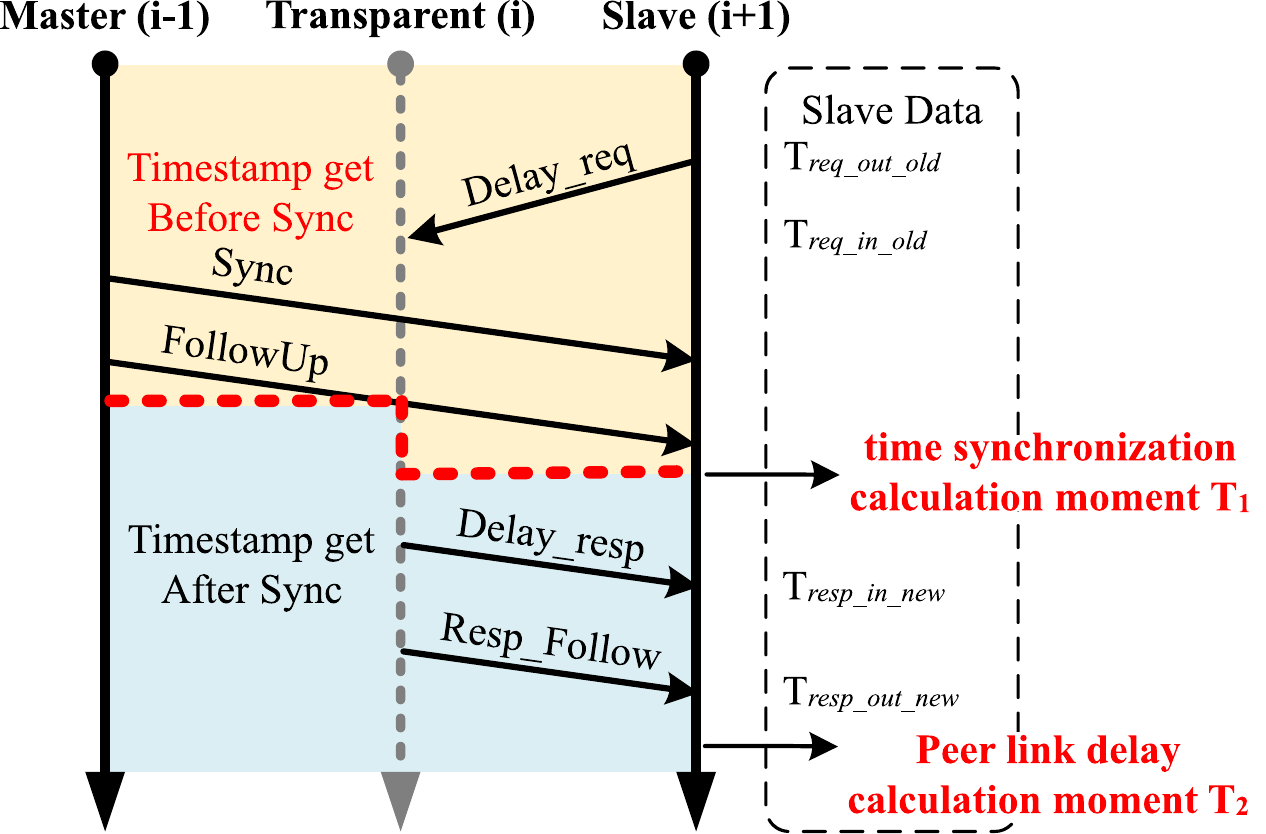}
    \caption{Synchronization Collision Scenario 1: \textcolor{black}{The transparent clock first transmitted the Sync message to the slave node and later responded to its peer delay measurement request}.}
    \label{fig:SyncCrashCondition1}
\end{figure}

According to Fig.~\ref{fig:SyncCrashCondition1}, at $T_\textit{req\_in}^{i}$, transparent clock $i$ receives $\textit{Dealy\_req}$ sent by its child node $i+1$ at $T_\textit{req\_out}^{i+1}$. However, before responding to the $\textit{Delay\_req}$, transparent clock $i$ receives $\textit{Sync}$ and $\textit{Follow Up}$ from its parent node $i-1$ and sends them to its child node. After completing the synchronization, transparent clock $i$ responds to the $\textit{Delay\_req}$ from child node $i+1$. Therefore, in the delay calculation of node $i+1$, timestamp $T_\textit{req\_out\_old}^{i+1}$ and $T_\textit{req\_in\_old}^{i+1}$ are generated before synchronization. Timestamp $T_\textit{resp\_out\_new}^{i}$ and $T_\textit{resp\_in\_new}^{i+1}$ are generated after synchronization.

Assuming that transparent clock $i$ has an offset value of time $\textit{Offset}_{i}$ after synchronization, the peer delay measurement error in child node $i+1$ due to time synchronization can be calculated as \eqref{eq:offset for pdelay1} based on \eqref{eq:peer delay measurement}. It can be observed that $\textit{Offset}_\textit{PeerDelay}$ is linearly related to the actual compensation during the delay calculation between master and slave. When a child node calculates peer delay, both the parent and child nodes are synchronized. We can compensate the timestamp generated before synchronization based on the \eqref{eq:offset for pdelay1}.
\begin{align}
    \textit{Offset}_{i}&=T_\textit{resp\_out\_new}^{i} - T_\textit{resp\_out\_old}^{i} \nonumber \\
    \textit{Offset}_\textit{PeerDelay}&=\frac{1}{2} \times \textit{rateRatio} \times \textit{Offset}_{i+1} \nonumber \\
    &-\frac{1}{2} \times \textit{Offset}_{i} \label{eq:offset for pdelay1}
\end{align}

\subsubsection{Synchronization Collision Scenario 2}
\begin{figure}[!t]
    \centering
    \includegraphics[width=\columnwidth]{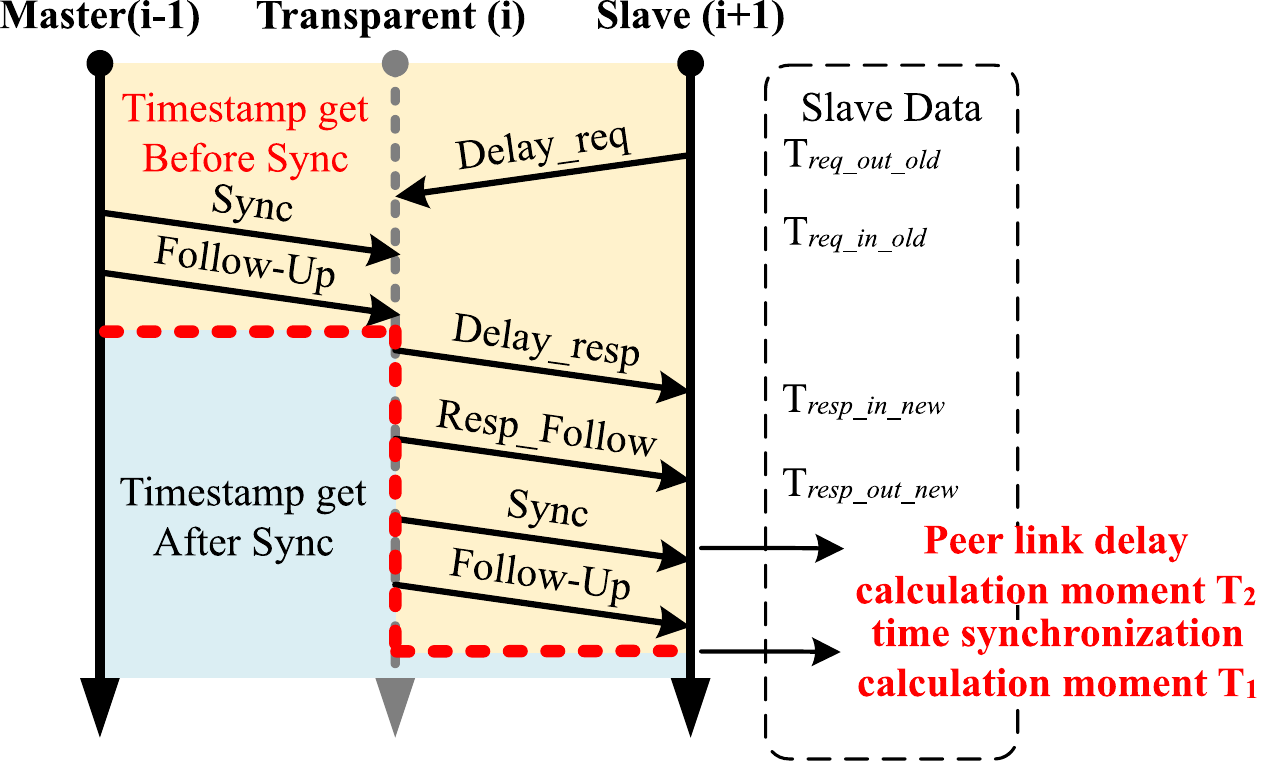}
    \caption{Synchronization Collision Scenario 2: \textcolor{black}{The transparent clock first responded to the peer delay measurement request and later transmitted the Sync message to its slave nodes}.}
    \label{fig:SyncCrashCondition2}
\end{figure}

Similar to scenario 1, child node $i+1$ sends $\textit{Delay\_req}$ to transparent clock $i$, and parent node $i-1$ sends $\textit{Sync}$ and $\textit{Follow Up}$ to transparent clock $i$ in sequence. The difference from scenario 1 is that transparent clock $i$ responds to the $\textit{Delay\_req}$ from child node $i+1$ first and then transfers the $\textit{Sync}$ to child node $i+1$. According to Fig.~\ref{fig:SyncCrashCondition2}, the sync error is determined by \eqref{eq:offser for pdelay2}. When a child node calculates peer delay, only the parent node is synchronized, and child nodes cannot compensate for timestamps.
\begin{align}
    \textit{Offset}_\textit{PeerDelay}&=rateRatio \times \textit{Offset}_{i+1} \nonumber \\
    &- \frac{1}{2} \times \textit{Offset}_{i} \label{eq:offser for pdelay2}
\end{align}

\subsubsection{\textcolor{black}{Dual Clock Counter Mode}}
The second scenario of the synchronization collision problem cannot be entirely resolved through software-level design. To address the time synchronization jitter caused by this problem, we reconfigured the nodes' clocks. \textcolor{black}{We introduced two counters for the PTP clock, one representing local time and the other global time. We named this mode "Dual Clock Counter Mode" to distinguish it from the "Single Clock Counter Mode", which uses the same clock counter to generate message timestamps and apply time synchronization offset.} Local time is influenced solely by the crystal oscillator, while global time is adjusted based on time synchronization protocol compensations. In our synchronization mechanism, we generate the synchronization timestamp based on local time and offset the global time based on the calculation. The same design principle is also applied in the time synchronization mechanism of TSN. We will verify the synchronization collision improvement mechanism's synchronization performance in Sec.~\ref{sec:Performance of Dual Clock Counter Mode}.

\subsection{Handling Sync Error in Re-transmission}
IEEE 1588 points out that the uncertainty in wireless communication implies that the peer delay between a mobile terminal and its parent node is less stable than in wired networks. The wireless channel's peer delay varies dynamically with the relative position of the mobile terminal and its parent node. IEEE 1588 recommends binding the synchronization and peer delay measurement in wireless channels. When a child node receives $\textit{Sync}$ and $\textit{Follow Up}$ message, It immediately initiates a peer delay measurement to obtain the latest peer delay. Subsequently, it calculates the synchronization offset.

\begin{figure}[!t]
    \centering
    \includegraphics[width=1\columnwidth]{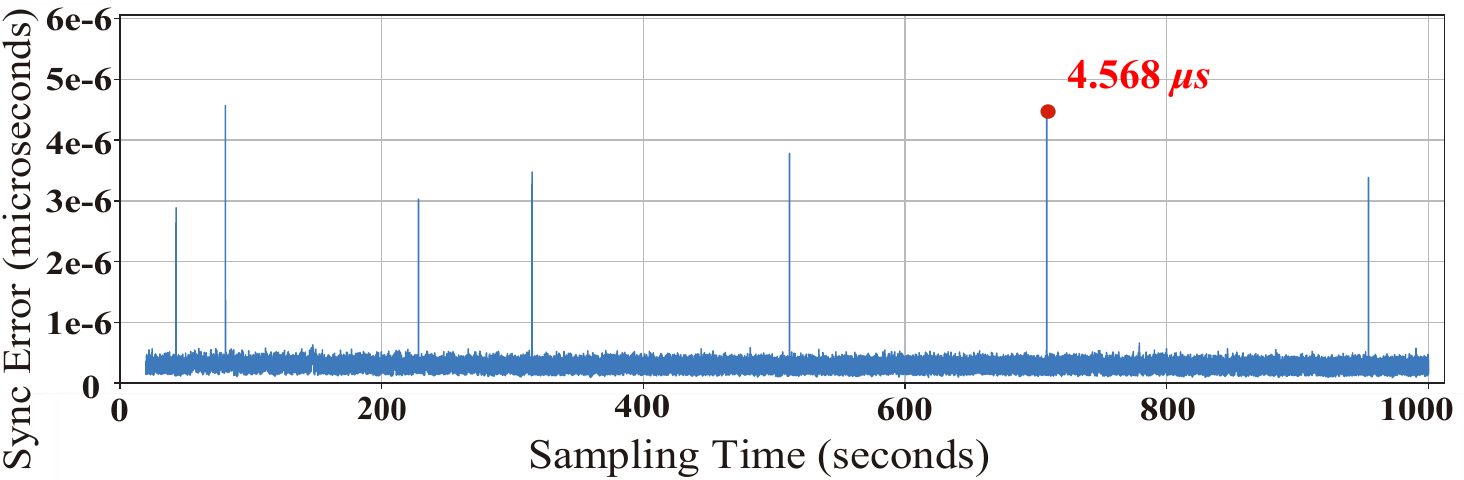}
    \caption{Sampling Result for Time Synchronization \textcolor{black}{when processing time synchronization messages using traditional HARQ re-transmission techniques in NG-RAN. $4.568\mu s$ is the maximum value of sync errors in the results, and each sync error persisted for some time before being eliminated.}}
    \label{fig:Harq Crash Result}
\end{figure}

For the acquisition of timestamps within the wireless network interfaces, we continue to utilize an interrupt-driven approach. We conducted point-to-point testing after implementing the time synchronization mechanism in the NG-RAN. We examined the synchronization performance under various wireless channel qualities. The influence of wireless channel quality on synchronization performance is significant. Fig.~\ref{fig:Harq Crash Result} shows the time synchronization performance with a Target Block Error Rate (BLER) of $0.001\%$.

\begin{figure}[!t]
    \centering
    \includegraphics[width=\columnwidth]{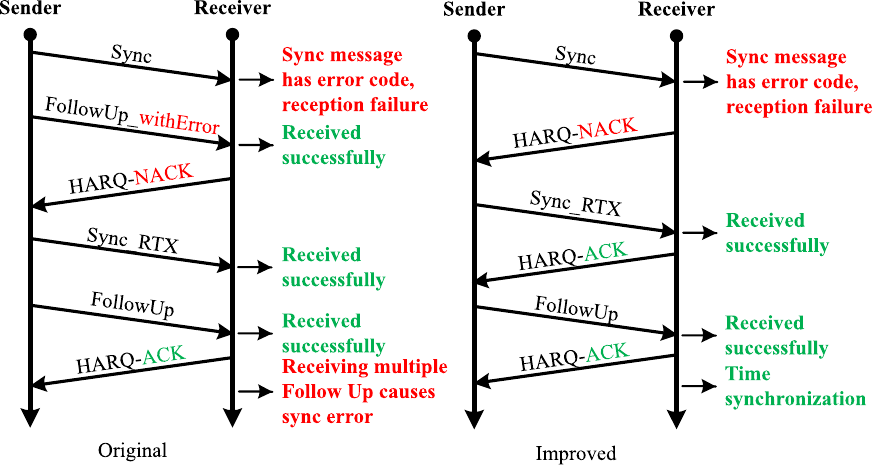}
    \caption{The Workflow of Hybrid Automatic Repeat Request (HARQ) \textcolor{black}{and the transmission flow of the synchronization message after using the improved mechanism}.}
    \label{fig:HARQ_error}
\end{figure}

In our analysis of the result, we observed that the HARQ processing mechanism in the 5G wireless network interface operates at the MAC Layer, while the interrupt generation mechanism is located between the MAC Layer and the Physical Layer. According to Fig.~\ref{fig:HARQ_error}, when the MAC Layer hands over the MAC frame to the physical layer for transmission, it does not immediately clear the backup of that MAC frame in the MAC Layer buffer. When the physical layer receives the $\textit{HARQ-ACK}$ signal from the receiver, it notifies the MAC Layer that the buffer for this frame can be released. However, if a $\textit{HARQ-NACK}$ signal is received, the physical layer will request the frame's information from the MAC Layer for re-transmission. In the event of a re-transmission, the same data packet is processed multiple times by the interrupt handler, causing multiple processing events by the upper-layer PTP process.

In a scenario involving only one re-transmission, when the $\textit{Sync}$ is first transmitted but lost, the interrupt handler reports the transmission timestamp and sends $\textit{Follow Up}$ with an incorrect timestamp. This leads to erroneous synchronization behavior. After the re-transmission, the $\textit{Follow UP}$ with the correct timestamp is discarded due to synchronization sequence ID mismatches (the upper-layer synchronization calculation process requires matching message ID to ensure that the right message information is used for calculations).

HARQ is vital for the stable operation of the 5G system. To mitigate the impact of HARQ on time synchronization, we analyzed the implementation details of HARQ and modified the time synchronization mechanism in the NG-RAN. Traditionally, when the interrupt mechanism detects data packet transmission and reports this behavior to the upper-layer PTP program, the upper-layer PTP program sends the $\textit{Follow UP}$ immediately. To address this, we introduced an interrupt mechanism in the wireless network interface's Medium Access Control (MAC) layer. After sending $\textit{Sync}$, the upper-layer PTP program will be waiting for a $\textit{HARQ-ACK}$ message from MAC before sending $\textit{Follow UP}$, illustrated in the right workflow of Fig.~\ref{fig:HARQ_error}. We will verify the synchronization performance of the improvement mechanism in Sec.~\ref{sec:perfromance of HARQ}.

\subsection{Improve Synchronization Precision in Mobility} \label{sec:path pelay}
In IEEE 1588, the mechanism of binding time synchronization and peer delay measurement has mitigated the impact of the uncertainty in wireless channels on synchronization performance. However, when nodes are in high-speed motion, the time interval between the occurrence of time synchronization and peer delay measurement introduces significant errors in the calculations. The binding behavior increases the occupancy of network bandwidth for time synchronization in the wireless channel. It changes from having two downstream data flows to four downstream data flows and one upstream data flow, increasing network bandwidth requirements for time synchronization by $150\%$. Each data stream in the wireless channel is at risk of transmission failure, and the increased number of interactions amplifies the probability of synchronization failure.

\begin{figure}[!t]
    \centering
    \includegraphics[width=\columnwidth]{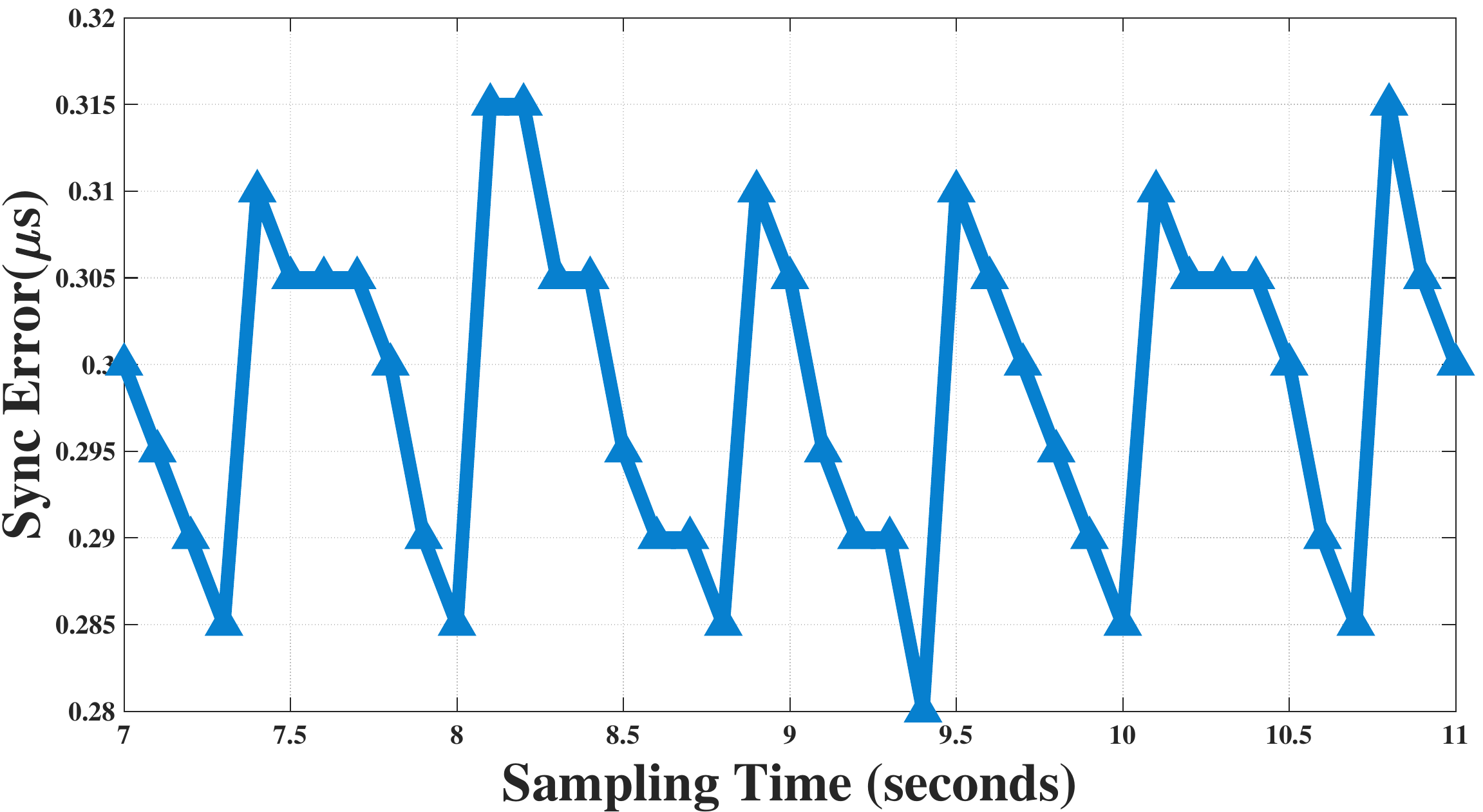}
    \caption{Sampling Result For Time Synchronization \textcolor{black}{when the UE moves uniformly away from the gNB along the X-axis direction at a velocity of $\textit{10 m/s}$}.}
    \label{fig:Motion_Analysis}
\end{figure}

To analyze the reasons behind the impact of UE mobility on time synchronization precision, we consider a UE moving uniformly away from the gNB along the X-axis direction at a velocity of $\textit{10 m/s}$. The time synchronization interval is 0.125s, and the peer delay measurement interval is 1s. For the sake of clarity, we focus on the sync error between the UE and the master clock gNB, as shown in Fig.~\ref{fig:Motion_Analysis}.

It can be observed that the sync error within the wireless channel experiences periodic variations. In the simulation, as the UE moves uniformly away from the gNB, this periodic sync error variation is related to the actual and measured values of the peer delay during synchronization. Let the actual peer delay be denoted as $\textit{Peerdelay}_\textit{reality}$, and the measured peer delay as $\textit{Peerdelay}_\textit{measure}$. The relationship between these two quantities after each peer delay measurement is given by \eqref{eq:path delay error}. In this example, the linear movement of the UE results in $\textit{d(t)}$ changing in accordance with the UE velocity and the peer delay measurement interval.
\begin{align}
    \textit{Peerdelay}_{\textit{reality}}=\textit{Peerdelay}_{\textit{measure}}+\textit{d(t)} \label{eq:path delay error}
\end{align}

In the time synchronization of the UE, the time offset calculation is based on $\textit{Peerdelay}_\textit{measure}$. According to \eqref{eq:sync}. The formula for calculating the actual offset $offset_{reality}$ is as follows.
\begin{align}
    \textit{offset}_\textit{measure}&=\text{T}_6-\text{T}_5-\textit{Peerdelay}_\textit{measure} \nonumber \\
    & -\text{CF}_\textit{sync}-\text{CF}_\textit{followUp} \label{eq:offset measure} \\
    \textit{offset}_\textit{reality}&=\text{T}_6-\text{T}_5-\textit{Peerdelay}_\textit{reality} \nonumber \\
    & -\text{CF}_\textit{sync}-\text{CF}_\textit{followUp} \label{eq:offset reality}
\end{align}

According to \eqref{eq:offset measure} and \eqref{eq:offset reality}, the sync offset is given by:
\begin{align}
    \textit{offset}_\textit{Measure}-\textit{offset}_\textit{Reality}= \textit{d(t)} \label{eq:offset error}
\end{align}

Our analysis was conducted without binding time synchronization and peer delay measurement. In this case, the change in measurement results caused by $\textit{d(t)}$ is approximately periodic in Fig.~\ref{fig:Motion_Analysis}, making it easier to understand. Through analysis, we know that the offset is directly proportional to the $\textit{d(t)}$. Therefore, when time synchronization and peer delay measurement are binding, based on the timestamp in Fig.~\ref{fig:P2PTransparentClock}, we can convert \eqref{eq:offset error} to \eqref{eq:offset error new}:
\begin{align}
    \textit{offset}_\textit{Measure} - \textit{offset}_\textit{Reality} = d \left(T_{9} \right) - d \left(T_{5} \right) \label{eq:offset error new}
\end{align}

According to \eqref{eq:offset error new}, it is evident that when the UE is moving, the interval between the peer delay measurement moment and the time synchronization moment leading to variations in peer delay will raise the sync error. Therefore, we propose an improved algorithm for using IEEE 1588 in wireless channels as shown in \eqref{eq:new peer delay}. 
\begin{align}
    \textit{PeerDelay} = T_6 - T_5 \label{eq:new peer delay}
\end{align}

During the network initialization phase, we restrained mobile terminals' movement. We employed a traditional time synchronization algorithm to establish high-precision time synchronization between the mobile terminals and their parent nodes. After waiting for the algorithm to converge, the clocks between the master and slave nodes are synchronized. Then, the mobile terminals calculate peer delay with \eqref{eq:new peer delay} based on Fig.~\ref{fig:P2PTransparentClock}.

\section{Evaluation and Validation} \label{sec:test and validate}
In this section, we evaluate and validate the improved time-synchronization mechanism for 5G+TSN. In the 5G and TSN integrated networking, we consider three possible scenarios for setting up the master clock in the time synchronization domain:
\begin{itemize}
    \item Utilizing nodes within the TSN (TSN master clock).
    \item Utilizing nodes within the 5G Core Network (gNB master clock).
    \item Utilizing UE within the NG-RAN (UE master clock).
\end{itemize}

\subsection{Experiment Setup}
OMNeT++ (Objective Modular Network Testbed in C++) is an object-oriented, modular, discrete event network simulation platform \cite{OMNeT}. INET (Internet Engineering Task Force Network Emulator) is an open-source software package developed based on OMNeT++. It also allows for the rapid design and validation of network protocols or models based on specifications \cite{INET}. Simu5G (Simulation Platform for 5G Networks) is a 5G simulation tool based on the OMNeT++ network simulation platform. It is suitable for simulating and analyzing the performance and behavior of 5G networks in various scenarios \cite{Simu5G}.

\subsubsection{5G+TSN Network Topology}
\begin{figure}[!t]
    \centering
    \includegraphics[width=\columnwidth]{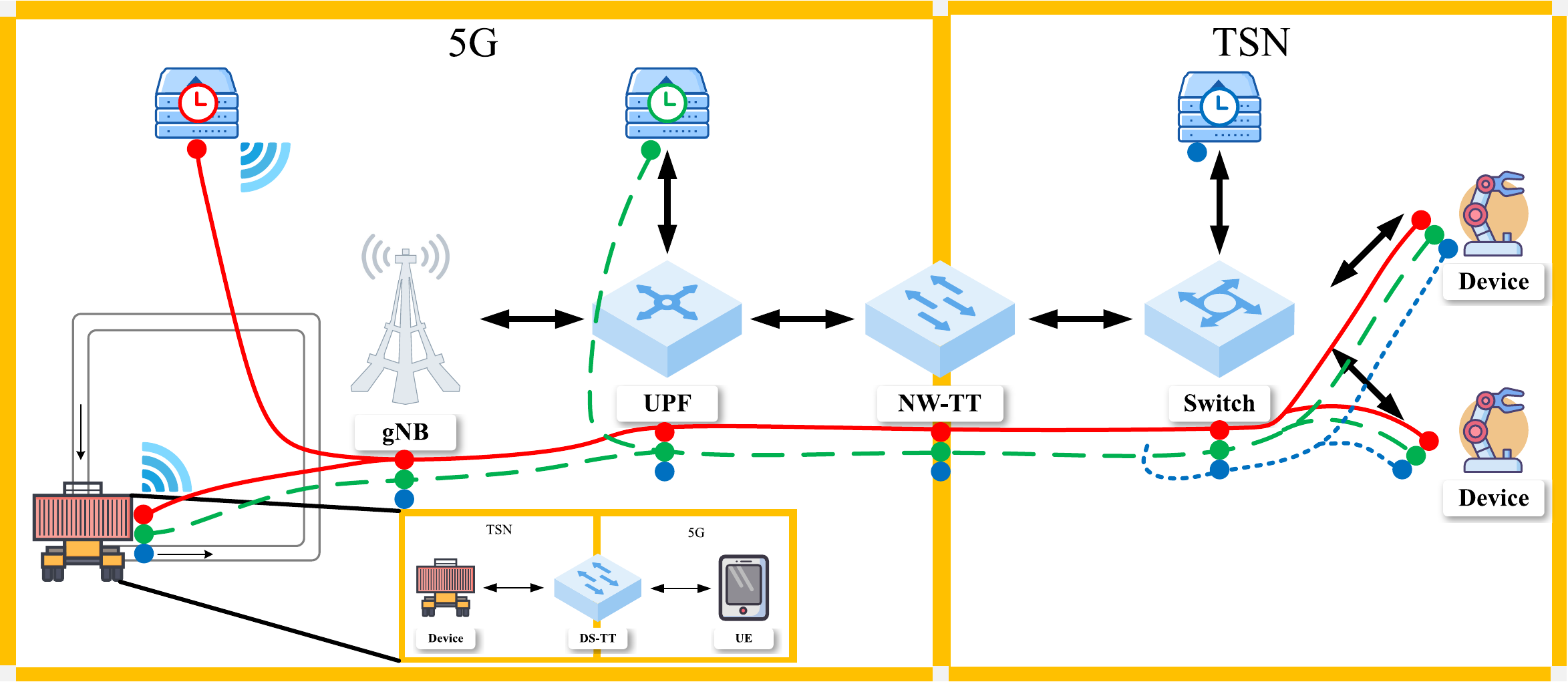}
    \caption{Simulation Verification Network Topology.}
    \label{fig:simulation network}
\end{figure}

We have built a 5G+TSN network topology in OMNeT++, as shown in Fig.~\ref{fig:simulation network}. The figure's red, green, and blue lines represent the path of time synchronization under the UE master clock, gNB master clock, and TSN master clock. In the network topology, UE performs rectangular motion around gNB according to the speed requirements of different experiments. Under the UE master clock, the clock source sends synchronization information to the network through a wireless connection.

\subsubsection{Configuration Parameters} \label{sec:simulation parameters}
In the simulation validation, we employed the user-recommended parameters from IEEE 802.1AS for time synchronization that performs time sync eight times per second and measures peer delay once per second for a wired TSN network. Additionally, we sampled the synchronization performance in the network 4000 times per second. Tab.~\ref{tab:Simulation Parameters} presents all the relevant parameters used in the simulation, with the 5G carrier frequency-related parameters based on the actual deployment parameters of the current communication operator for the 3.5 GHz spectrum.

\begin{table}[!t]
    \centering
    \caption{Simulation Parameters}
    \begin{tabular}{ll}
        \toprule
        \textbf{parameter} & \textbf{value} \\
        \midrule
        Bandwidth in TSN & 1Gbps \\ 
        Bandwidth in 5G Core Network & 10Gbps \\ 
        Clock frequency & 200MHz \\ 
        Clock drift for master & [-5ppm, 5ppm] \\ 
        Clock drift for others & [-10ppm, 10ppm] \\ 
        Sync interval & 0.125s \\ 
        Peer delay measurement interval & 1s \\ 
        Sync performance sampling frequency & 4KHz \\ 
        Carrier frequency & 3.5GHz \\ 
        Sub-carrier interval & 60KHz \\ 
        Number of Resource Blocks & 135 \\ 
        gNB/UE wireless transmission power & 46dBm/26dBm \\ 
        gNB/UE antenna height & 25m/1.5m \\ 
        Target BLER & 0.0001\% \\
        \bottomrule
    \end{tabular}
    \label{tab:Simulation Parameters}
\end{table}

\subsubsection{Sampling of time synchronization}
In simulation, we sample the time synchronization by recording the maximum clock offset among all synchronized global clocks. We collected the sampling result's mean, variance, and maximum error to demonstrate the time synchronization performance.

During time synchronization, network nodes are often affected by various complex factors, such as network congestion. As a result, nodes can not be synchronized for a long time. We use the concept of loss sync proportion to evaluate such impacts. In simulation, loss sync refers to nodes not synchronized in three continuous sync intervals.

\begin{figure}[!t]
    \centering
    \includegraphics[width=\columnwidth]{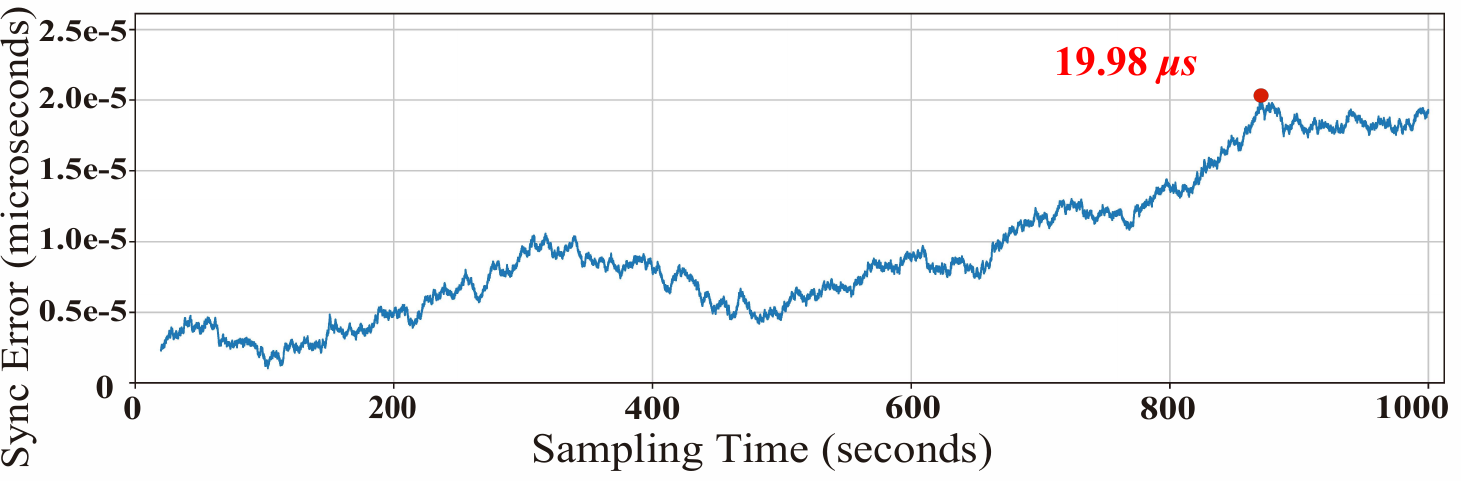}
    \caption{Sampling Results Without Time Synchronization.}
    \label{fig:Controlled experiment}
\end{figure}

To show our experimental effect more accurately, the figure Fig.~\ref{fig:Controlled experiment} initialize the results without time synchronization for the 5G+TSN network in Fig.~\ref{fig:simulation network} and parameters in Tab.~\ref{tab:Simulation Parameters}.

\subsection{Performance of time-synchronization improvement mechanisms}
\subsubsection{Performance After Handling gNB Competition} \label{sec:Performance of Dual Clock Counter Mode}
\begin{figure}[!t]
    \centering
    \includegraphics[width=\columnwidth]{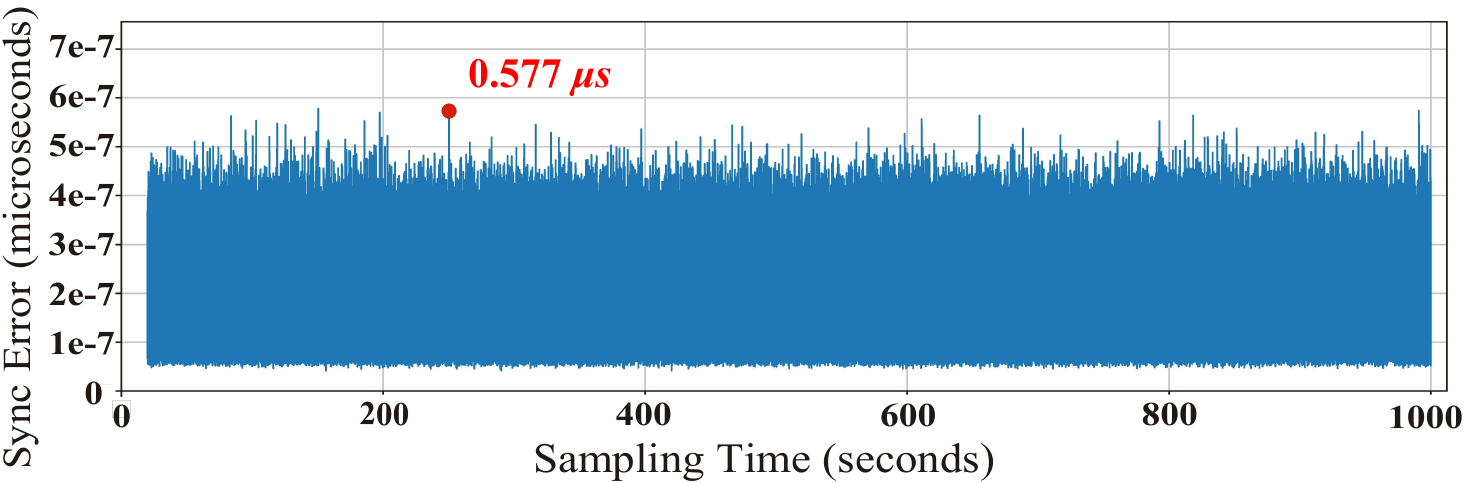}
    \caption{Sampling Result For Time Synchronization with Using Dual Clock Counter.}
    \label{fig:After dual counter}
\end{figure}

In the simulation, we sample the time synchronization in dual clock counter mode using the same parameter settings as in Sec.~\ref{sec:Dual Clock Counter Mode}. Under the same parameter settings, \textcolor{black}{the dual clock counter mode} effectively controls the sync error within $1\mu s$. In the simulation with a duration of 1000s, there was no abnormal synchronization jitter similar to that in single clock mode, as shown in Fig.~\ref{fig:After dual counter}.

\subsubsection{Performance After Handling Re-transmission} \label{sec:perfromance of HARQ}

In the proposed design to eliminate the impact of HARQ on time synchronization by adding MAC layer interruption, we mentioned that the increment of time interval between the $\textit{Sync}$ and $\textit{Follow Up}$ will be extending the overall synchronization time of the system. 

\begin{figure}[tbp]
    \centering
    \includegraphics[width=\columnwidth]{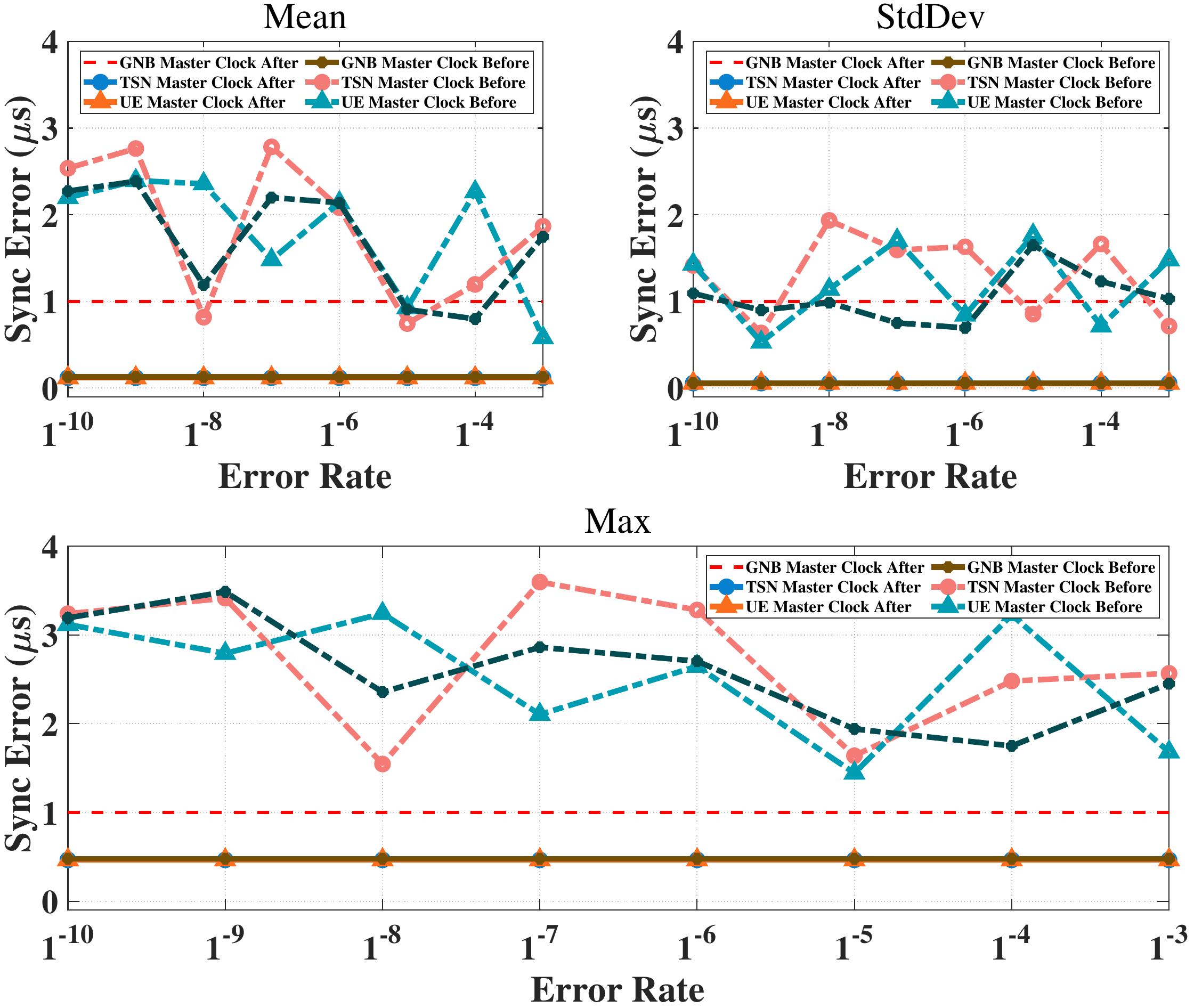}
    \caption{Synchronization Performance With Different Target BLER. \textcolor{black}{"Before" indicates the case of synchronization error under different Target BLERs without using the improvement. "After" indicates the effect of our proposed improvement.}}
    \label{fig:Accuracy}
\end{figure}

In synchronization, the worse the wireless channel quality, the longer the overall synchronization time. We tested the impact of different wireless channel error rates on time synchronization. We tested the time synchronization performance without using MAC layer interruption. We demonstrate that this impact of increasing sync time is limited for synchronization performance in Fig.~\ref{fig:Accuracy}. The results show that HARQ re-transmission events under different Target BLERs do not cause any more synchronization errors in the wireless channel.

\subsection{Parameter comparison and robustness analysis}
\subsubsection{Synchronization Interval}
\begin{figure}[!t]
    \centering
    \includegraphics[width=\columnwidth]{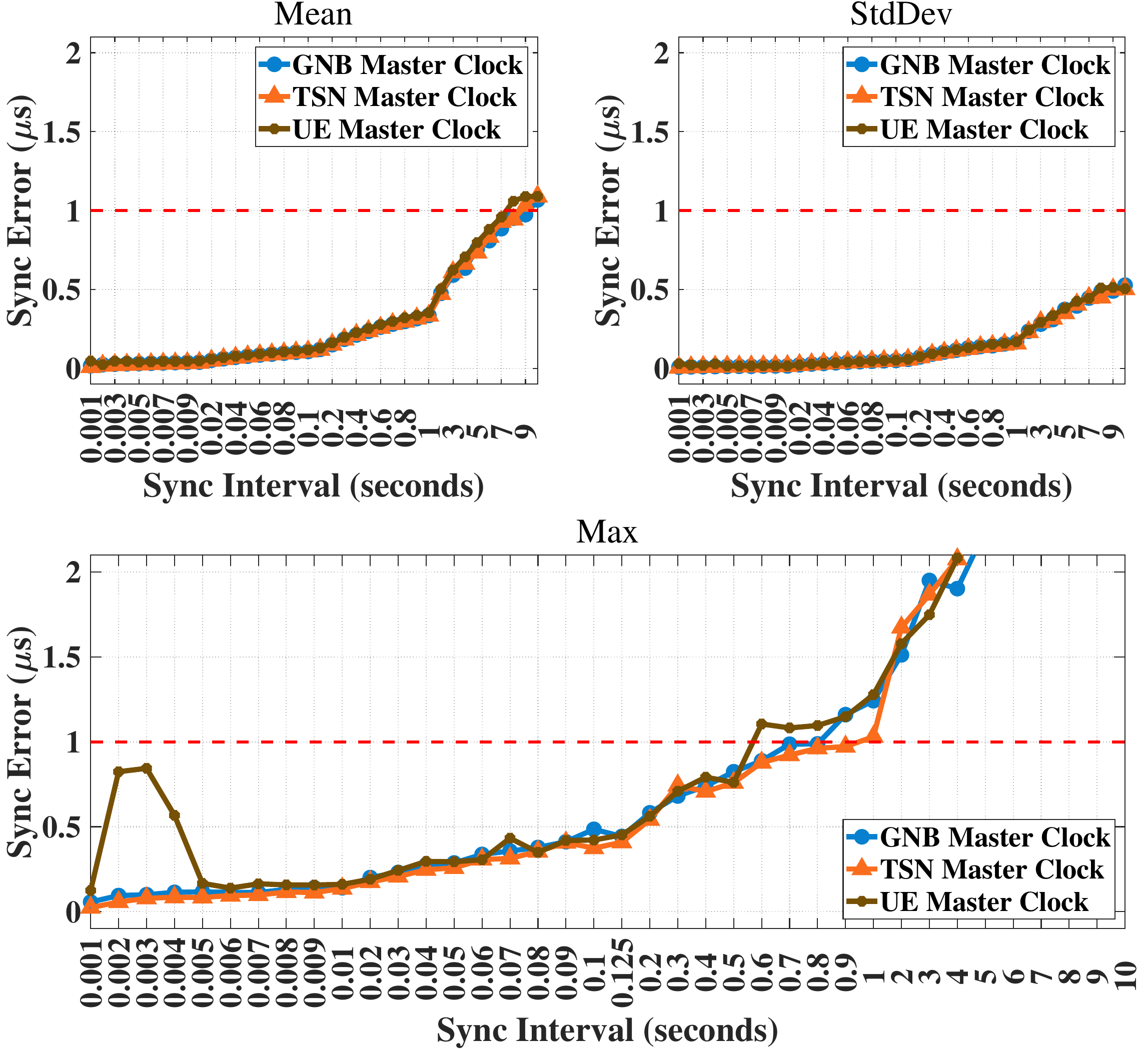}
    \caption{Synchronization Performance with Different Sync Intervals.}
    \label{fig:Sync Interval}
\end{figure}

In devices with local clocks based on oscillators, time deviations accumulate over time. The sync interval directly affects synchronization performance. Moreover, the sync interval determines the network bandwidth usage by synchronization messages. To verify whether this model meets different time synchronization requirements in various networks, we sample time synchronization with different time sync intervals as shown in Fig.~\ref{fig:Sync Interval}.

According to the result, time synchronization performance decreases as the time synchronization interval increases. The time synchronization performance details for a sync interval of 0.125s are shown in Tab.\ref{tab:sync performance in 0.125s}. 

\begin{table}[!t]
    \centering
    \caption{Synchronization Performance at Sync Interval of 0.125s.}
    \setlength{\tabcolsep}{5mm}
    \begin{tabular}{cccc}
        \toprule
        \textbf{Master Clock} & \textbf{Mean} & \textbf{StdDev} & \textbf{Max} \\
        \midrule
        TSN & 0.118$\mu s$  & 0.055$\mu s$  & 0.409$\mu s$   \\
        gNB & 0.120$\mu s$  & 0.057$\mu s$  & 0.443$\mu s$$\mu s$   \\
        UE & 0.128$\mu s$  & 0.057$\mu s$  & 0.452$\mu s$   \\
        \bottomrule
    \end{tabular}
    \label{tab:sync performance in 0.125s}
\end{table}

\begin{figure}[!t]
    \centering
    \includegraphics[width=\columnwidth]{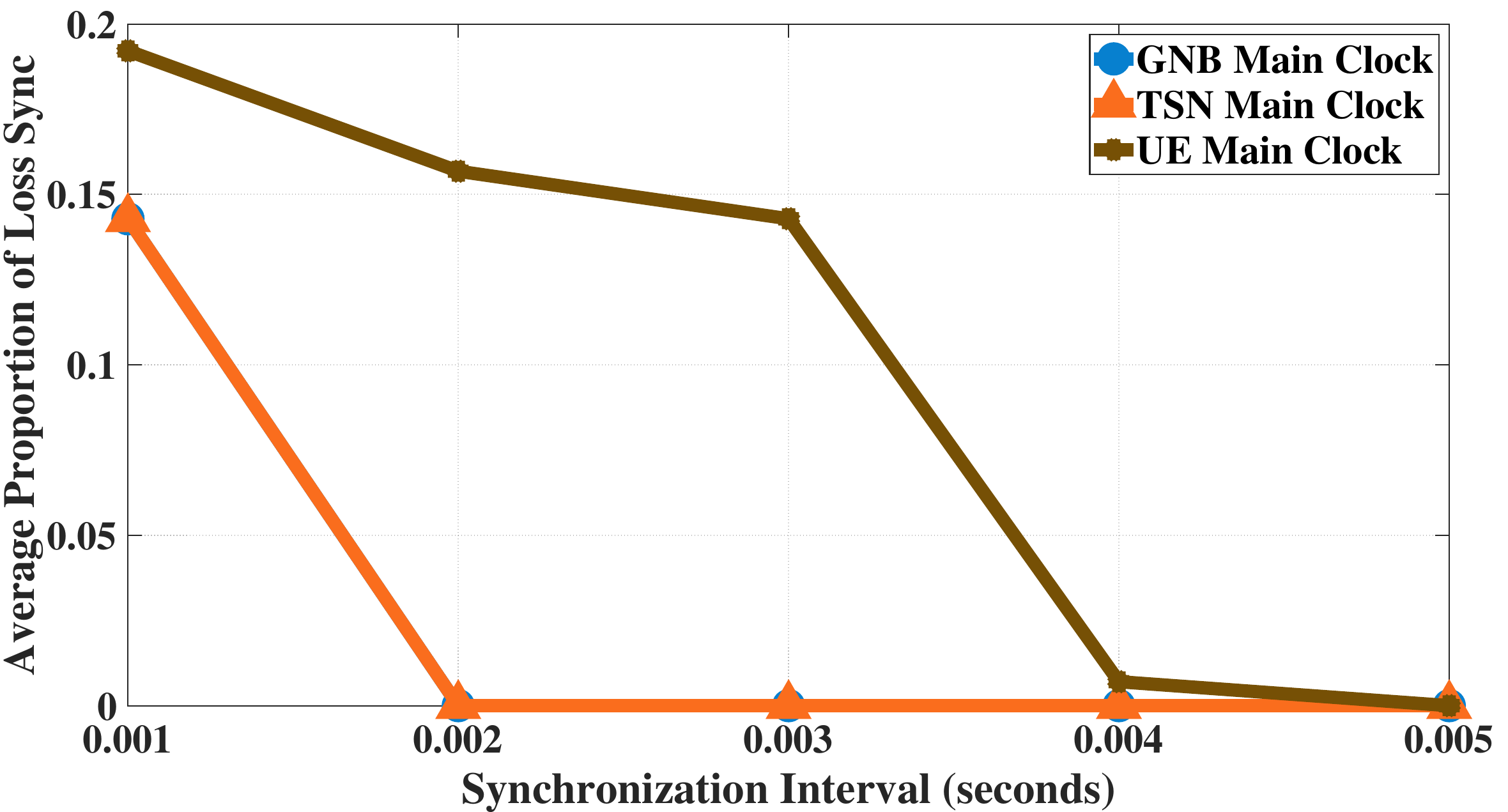}
    \caption{Loss Sync Proportion with Different Sync Intervals.}
    \label{fig:Loss Sync rate in interval}
\end{figure}

However, none of the three master clocks achieved better synchronization performance at sync intervals below 0.005s. Fig.~\ref{fig:Loss Sync rate in interval} illustrates the node's loss sync proportion at different sync intervals. The loss sync proportion is zero when the sync interval exceeds 0.005s. Under the same sync interval, the loss sync proportion of the UE master clock is always worse than the TSN master clock and gNB master clock.

As previously proposed to address sync errors caused by HARQ, we increased the interval between the $\textit{Sync}$ message and $\textit{follow Up}$ message. When this interval is bigger than the sync interval, the reception of a new $\textit{Sync}$ message during the waiting interval for the $\textit{follow Up}$ message leads to a sync error. Simultaneously, when the UE master clock fails to synchronize with its gNB, it triggers a chain reaction, causing all nodes in the wired network to lose synchronization. 

Our experimental results strongly discourage using UE master clocks for time synchronization. And under the precision requirement of $1\mu s$, the time sync interval should not exceed 0.5s.

\subsubsection{Network Diameter}
\begin{figure}[!t]
    \centering   
    \includegraphics[width=\columnwidth]{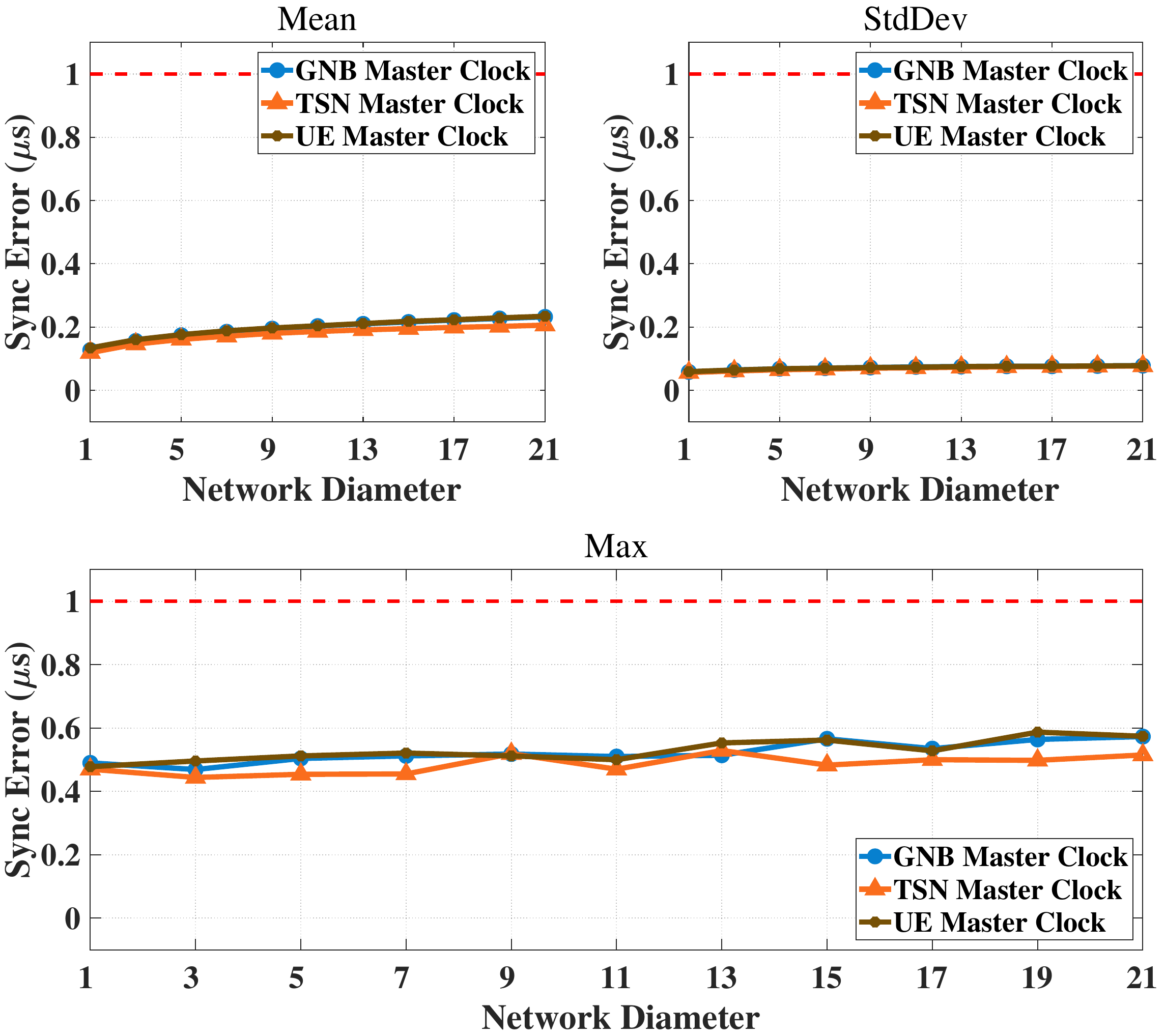}
    \caption{Synchronization Performance with Different Network Diameter.}
    \label{fig:Hop}
\end{figure}

In a time synchronization domain employing the Transparent Clock mechanism, the behavior of transparent clocks will impact the synchronization precision. According to the PTP definition of transparent clock, errors caused by the propagation delay accumulate with the distance of packet transmission. With a time sync interval of 0.125s, We sample time synchronization for network diameters ranging from 5 to 22 hops as shown in Fig.\ref{fig:Hop}.

The result demonstrates that synchronization precision gradually decreases as the network diameter increases. Furthermore, there was no loss of synchronized nodes in the network at any point. 

Under the precision requirement of $1\mu s$, the time synchronization mechanism is suitable for industrial environments with a maximum network diameter not exceeding 20 hops.

\subsubsection{UE Mobility}
Simu5G doesn't provide a specific implementation for the propagation delay in the wireless channel. When the UE is moving, the dynamic changes in propagation delay can affect synchronization performance to some extent and cannot be ignored. Therefore, we modified the relevant source code in Simu5G \cite{Bazion2018D}.

Assuming UE's instantaneous motion vector relative to the gNB is  $\overrightarrow{V_\textit{relative}}$, the component of the UE's instantaneous motion vector relative to the gNB along the speed of light is denoted as $\overrightarrow{V_{\textit{relative}\_\overrightarrow{C}}}$. The relative distance between the UE and gNB is $\textit{distance}$.
\begin{align}
    \overrightarrow{V_\textit{relative}}&=\overrightarrow{V_\textit{UE}}-\overrightarrow{V_\textit{gNB}} \nonumber \\
    delay&=\frac{\textit{distance}}{\left| \overrightarrow{C}+\overrightarrow{V_{\textit{relative}\_\overrightarrow{C}}} \right|} \label{eq:propagation delay}
\end{align}

\begin{figure}[!t]
    \centering
    \includegraphics[width=\columnwidth]{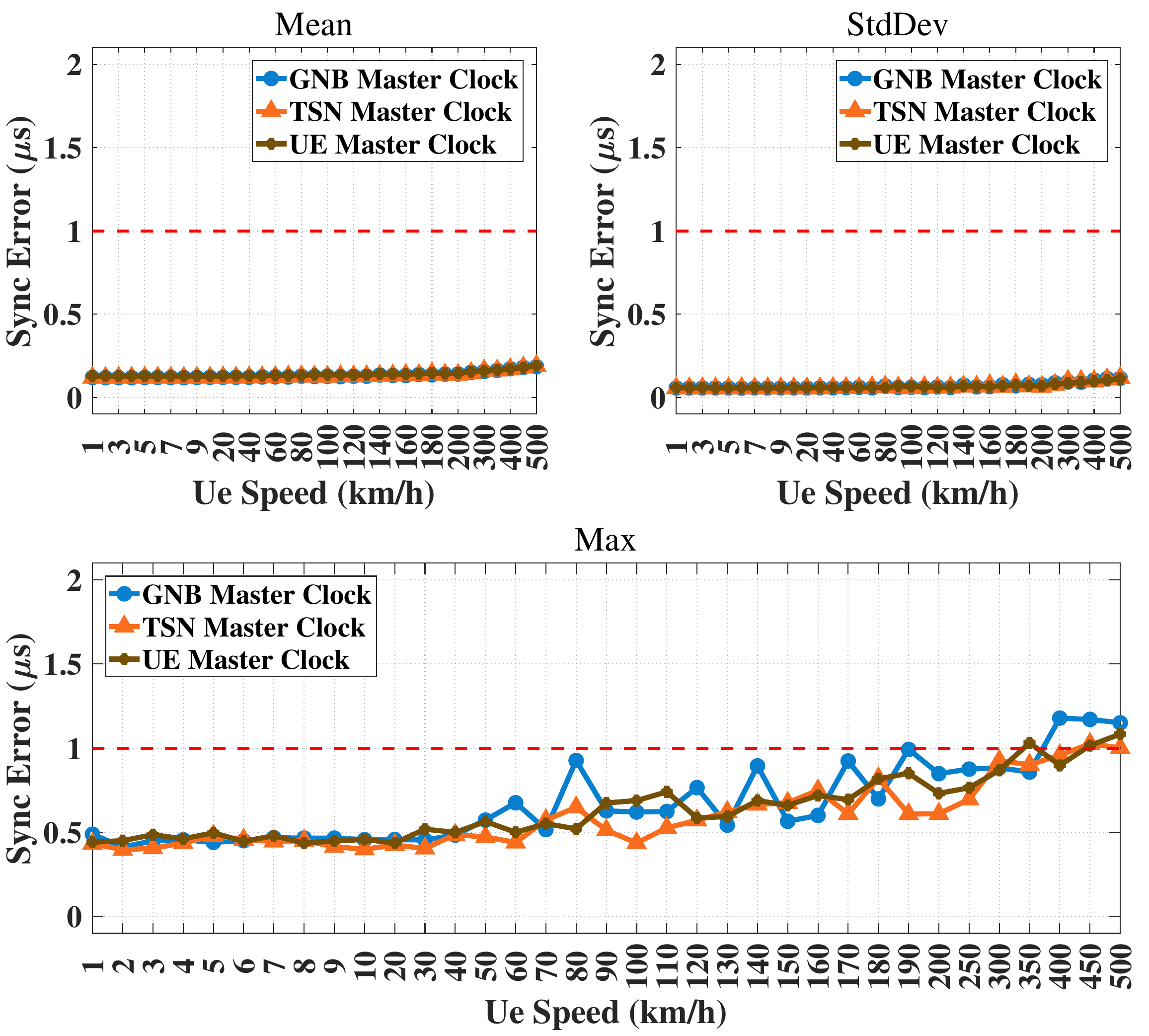}
    \caption{Synchronization Performance with Different UE Speed.}
    \label{fig:Moving}
\end{figure}

An essential feature of the 5G and TSN integrated networking is the enhancement of flexibility and flexible layout capabilities through wireless connections. With a network diameter of 5 hops and a time sync interval of 0.125s, We sample the time synchronization performance with UE movement speed in the range of 1km/h to 500km/h as shown in Fig.~\ref{fig:Moving}. 

Experimental results indicate that as the UE's movement speed increases, time synchronization precision decreases, and there is a certain degree of stochastic fluctuation. Our analysis in Sec.~\ref{sec:path pelay} demonstrates the limitations of IEEE 1588 based on link delay measurement at extremely high speed.

Under the precision requirement of $1\mu s$, the time synchronization mechanism is suitable for an industrial environment with the maximum speed of the UE does not exceed $200 km/h$.

\subsubsection{Network Load Rate}
\begin{figure}[!t]
    \centering
    \includegraphics[width=\columnwidth]{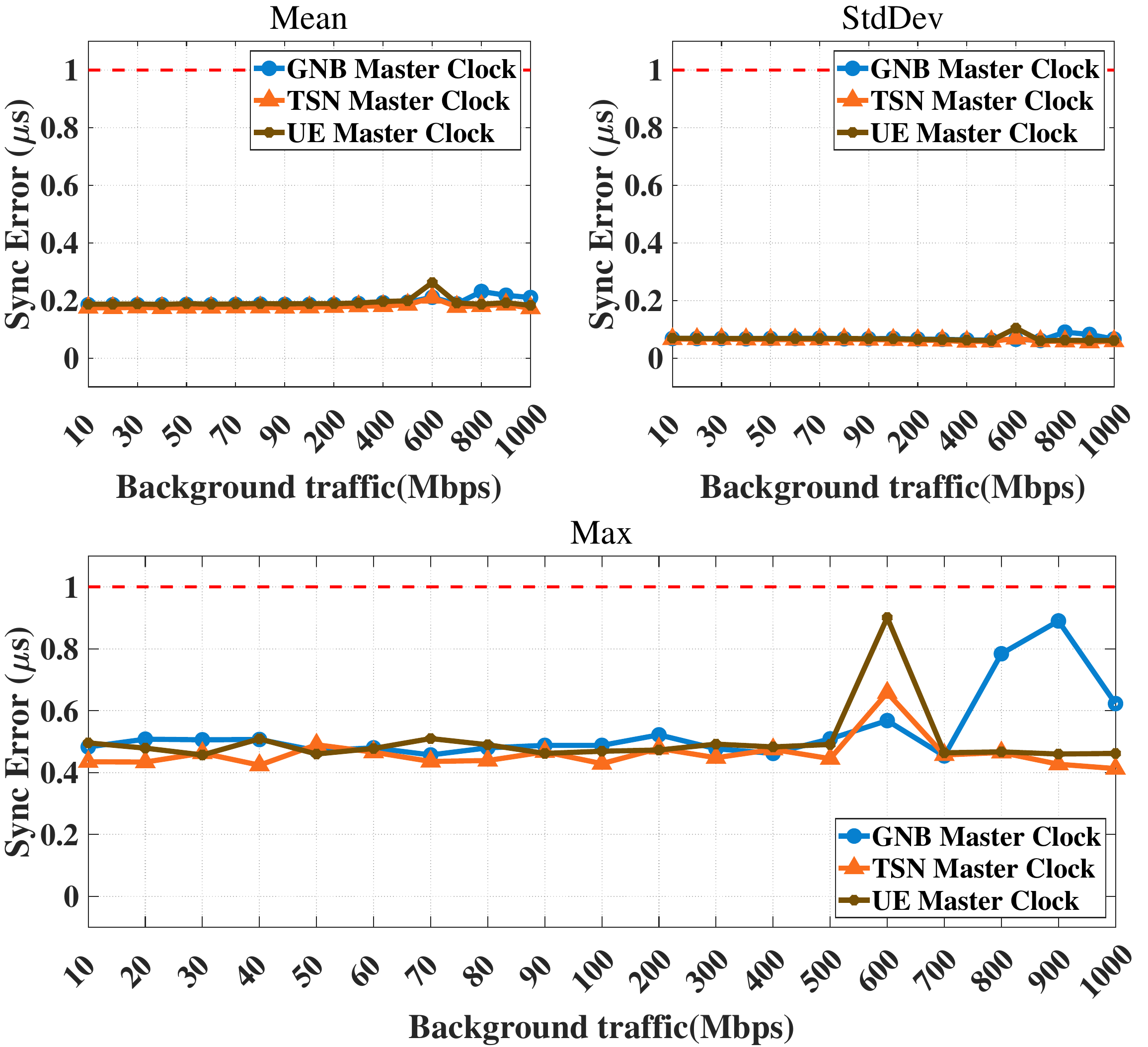}
    \caption{Synchronization Performance with Different Network Load Rate.}
    \label{fig:Flow}
\end{figure}

The network load rate represents the current network's bandwidth usage proportion to the total available network bandwidth. The characteristics of the PTP protocol result in synchronization precision being affected under high network loads. We sample the synchronization performance under network load rates from 10Mbps to 1Gbps, as shown in Fig.~\ref{fig:Flow}. 

\begin{figure}[!t]
    \centering
    \includegraphics[width=\columnwidth]{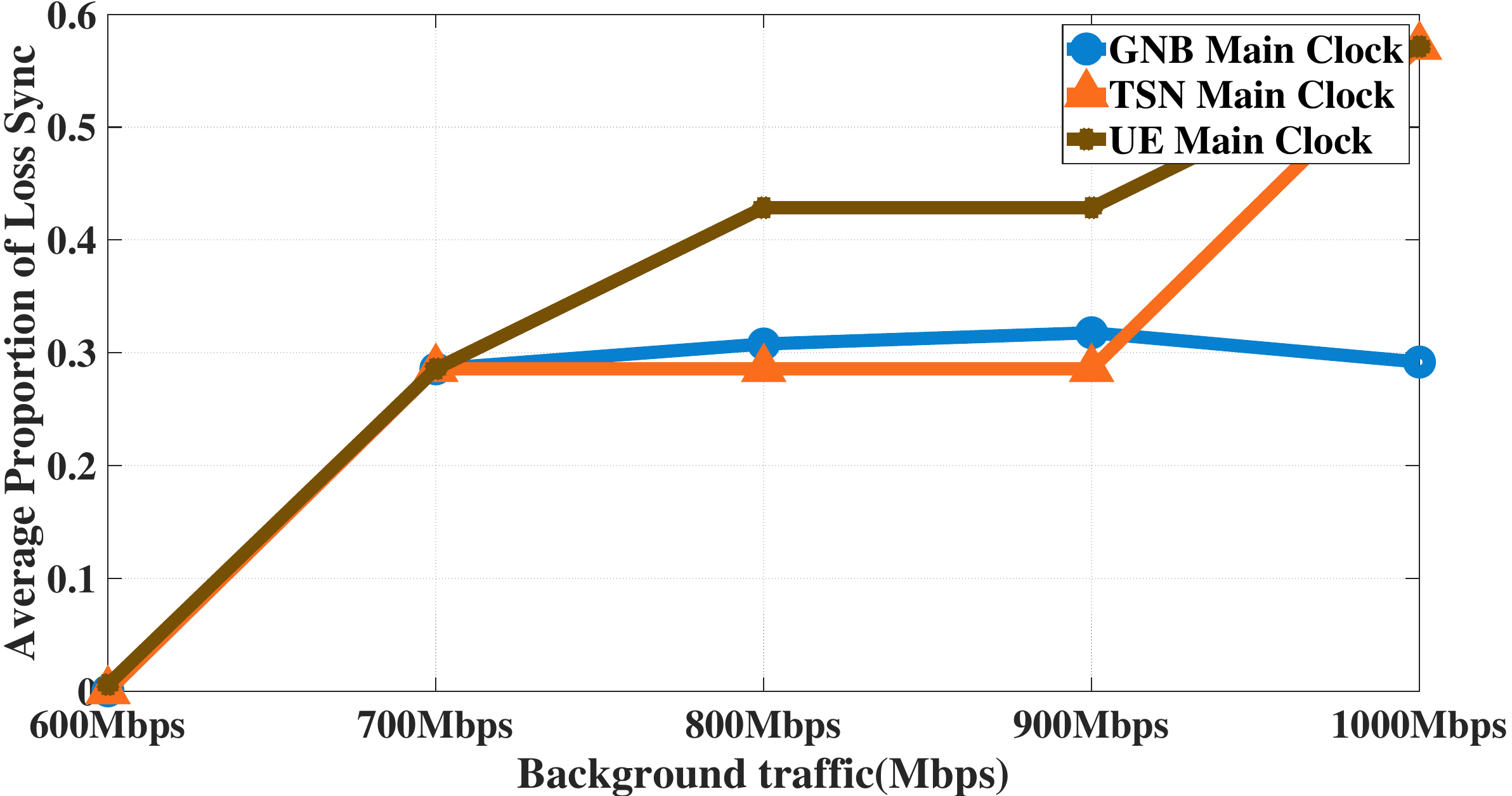}
    \caption{Loss Sync Proportion with Different Network Load Rate.}
    \label{fig:loss sync flow}
\end{figure}

The result indicates that when the network load rate is below 600Mbps, the impact of network load on synchronization precision is relatively tiny. However, when the network load rate exceeds 600Mbps, network congestion leads to increased delay and even packet loss in time synchronization messages, resulting in an exponential degradation of synchronization precision. At the same time, there was a significant occurrence of node loss synchronization in the network when the network load rate exceeds 600Mbps in Fig.\ref{fig:loss sync flow}.

Under the precision requirement of $1\mu s$, the time synchronization mechanism is suitable for an industrial environment with the maximum network load rate not exceeding $60\%$.

\subsection{Case Study On Industrial Environments}
\begin{figure}[!t]
    \centering
    \includegraphics[width=0.9\columnwidth]{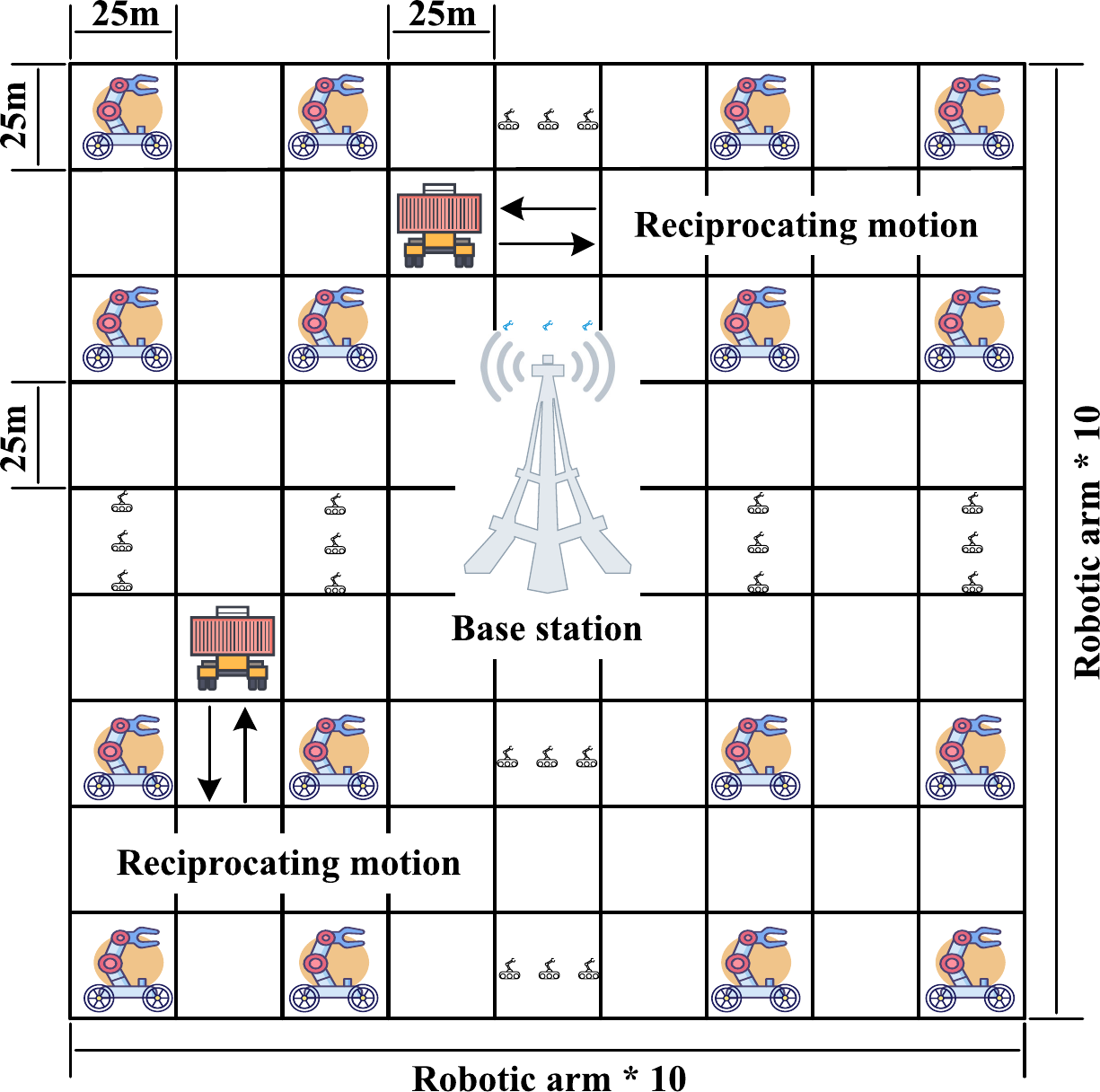}
    \caption{Simulation Network for Industrial Environment Experiment.}
    \label{fig:Working_environment}
\end{figure}

A large factory floor size of 500m $\times$ 500m is considered for the simulation layout wherein there is a mobile robot every 50 with 100 robots in total \cite{Kho20195G}. Each robot moves on a 25m $\times$ 25m scale based on their work tasks. Simultaneously, between any two mobile robots, one robot moves horizontally, and the other moves vertically, transporting material between different mobile robots. There are nine horizontally and nine vertically moving robots, as shown in Fig.\ref{fig:Working_environment}. 

\begin{figure}[!t]
    \centering
    \includegraphics[width=\columnwidth]{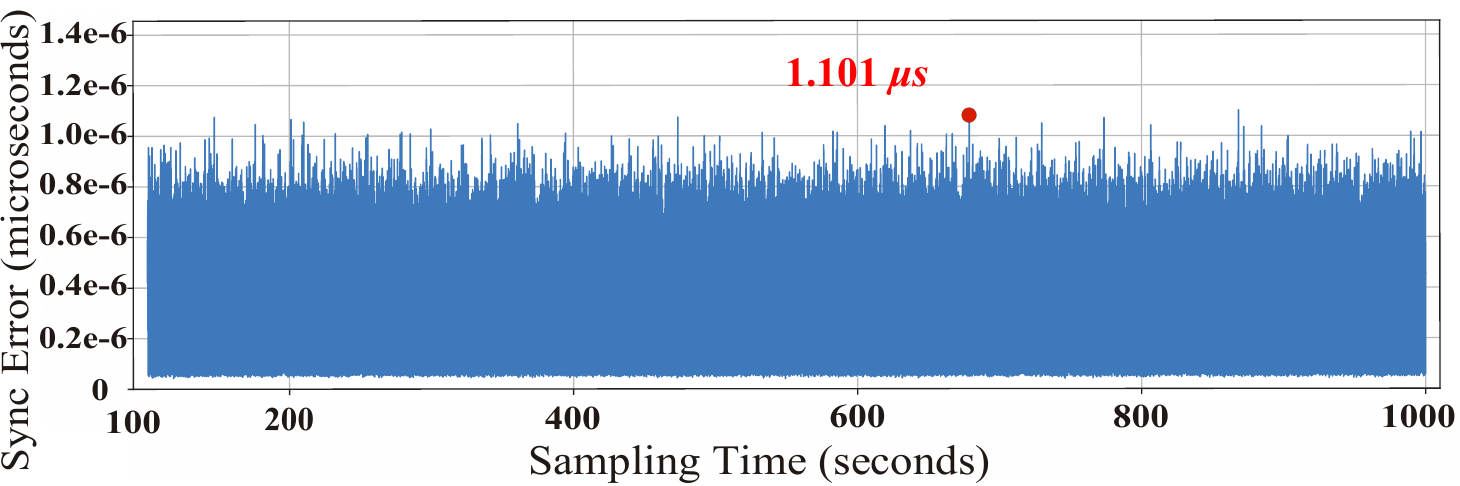}
    \caption{Sampling Result for Industrial Environment with TSN Master Clock.}
    \label{fig:Working environment sync result TSN}
\end{figure}

\begin{figure}[!t]
    \centering
    \includegraphics[width=\columnwidth]{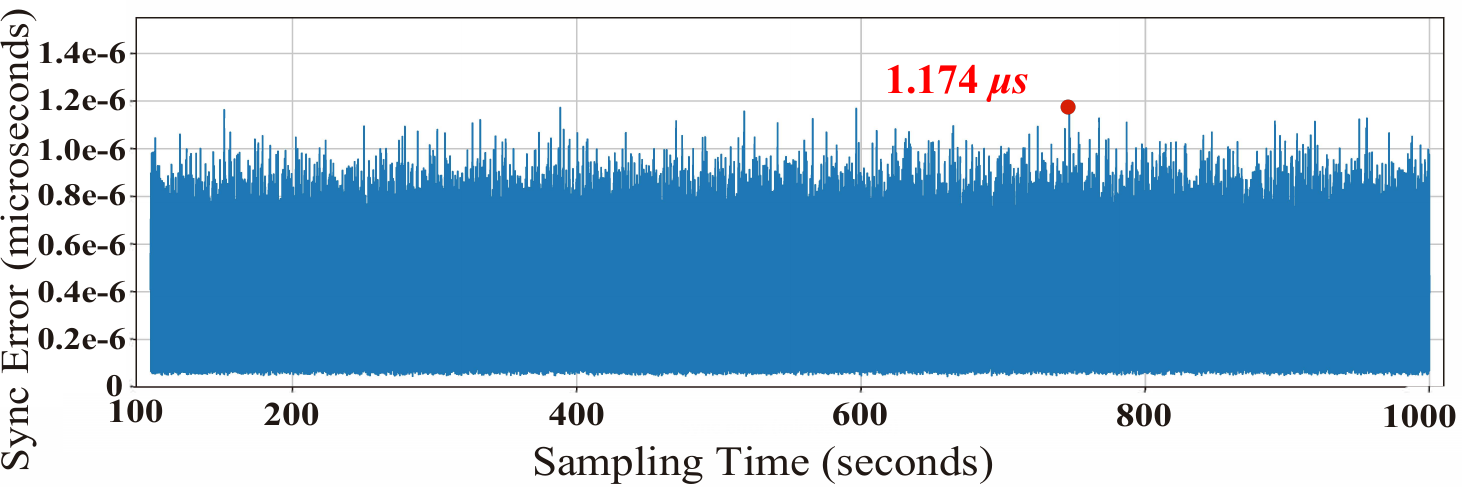}
    \caption{Sampling Result for Industrial Environment with gNB Master Clock.}
    \label{fig:Working environment sync result gNB}
\end{figure}

In previous robustness analyses, deploying the master clock in the wireless channel, namely UE as the master, was extremely not recommended since it would maximize the impact of the complexity and uncertainty of the wireless channel on time synchronization. In industrial environment testing, we no longer simulate the UE master clock. Based on the analysis results mentioned earlier, and the wireless industrial environment test cases proposed in \cite{Ati2022W}, we set the network simulation parameters as shown in Tab.~\ref{tab:Simulation Parameters For Industrial Environmental Testing} and basic parameters in Tab.~\ref{tab:Simulation Parameters}. Fig.\ref{fig:Working environment sync result TSN} and Fig.\ref{fig:Working environment sync result gNB} show the sync error sampling results for the network after running for 1000 seconds. We achieve a precision of nearly 1 microsecond with interoperability between 5G nodes and TSN nodes in industrial environments.

\begin{table}[!t]
    \centering
    \caption{Simulation Parameters For Industrial Environment}
    \begin{tabular}{ll}
        \toprule
        \textbf{parameter} & \textbf{value} \\
        \midrule
        Periodic flows & 100Mbps \\
        Sudden flows & Random in [0, 200] Mbps \\
        UE Speed & Random in [5, 10] m/s \\
        \bottomrule
    \end{tabular}
    \label{tab:Simulation Parameters For Industrial Environmental Testing}
\end{table}

\section{Conclusion} \label{sec:conclusion}
High-precision time synchronization serves as the foundation for achieving deterministic transmission in the integrated networking of 5G and TSN systems. Based on the IEEE 1588 and its variant IEEE 802.1AS, we improved the time synchronization mechanism for 5G and TSN integrated networking with interoperability. We systematically studied the time synchronization performance on the OMNeT++ network simulation platform using the INET and Simu5G network simulation framework. We summarize the suitable parameters allowed for time synchronization to achieve the precision requirement of $1\mu s$ for the industrial environment in Tab.~\ref{tab:Maximum parameters for sync}. Simulation results indicate that the proposed time synchronization mechanism can achieve a synchronization performance of no more than 1 microsecond with the interoperability of 5G and TSN nodes in industrial environments.

\begin{table}[!t]
    \centering
    \caption{Maximum Available Parameter for Time Synchronization}
    \begin{tabular}{l c c c}
        \toprule
        ~ & \multicolumn{3}{c}{Master Clock} \\
        ~ & \textbf{TSN} & \textbf{gNB} & \textbf{UE} \\
        \midrule
        Sync Interval (seconds) & 0.003-0.5 & 0.003-0.5 & 0.005-0.5 \\
        Network Diameter (hops) & 20 & 20 & 20 \\
        Working Speed (km/h) & 200 & 200 & 200 \\
        Network Load Rate & 50\% & 50\% & 50\% \\
        \bottomrule
    \end{tabular}
    \label{tab:Maximum parameters for sync}
\end{table}

\bibliographystyle{IEEEtran}
\bibliography{reference}


 





\end{document}